\documentclass[lettersize,journal,twoside]{IEEEtran}

\usepackage{amsmath,amsfonts}
\usepackage{algorithmic}
\usepackage{algorithm}
\usepackage{array}
\usepackage[caption=false,font=normalsize,labelfont=sf,textfont=sf]{subfig}
\usepackage{textcomp}
\usepackage{stfloats}
\usepackage{url}
\usepackage{verbatim}
\usepackage{graphicx}
\usepackage{cite}
\usepackage[colorlinks,linkcolor=blue,citecolor=blue]{hyperref}
\usepackage{caption}
\usepackage{float} 
\usepackage{subcaption}
\usepackage{booktabs}
\usepackage{multirow}

\usepackage{orcidlink}

\begin{document}

\title{Next-Gen Computing Systems with Compute Express Link: a Comprehensive Survey}

\author{Chen~Chen\orcidlink{0009-0003-1940-3037},
        Xinkui~Zhao\orcidlink{0000-0002-1115-5652},
        Guanjie~Cheng\orcidlink{0000-0003-2080-3903},
        Yuesheng~Xu\orcidlink{0000-0001-7210-0543},
        Shuiguang~Deng\orcidlink{0000-0001-5015-6095},
        and Jianwei~Yin\orcidlink{0000-0003-4703-7348}
\thanks{\textit{(Corresponding author: Xinkui Zhao.)}}

\thanks{Chen Chen, Xinkui Zhao, Guanjie Cheng, Shuiguang Deng, and Jianwei Yin are with the College of Computer Science and Technology, Zhejiang University, Hangzhou, Zhejiang, 310027, China (e-mail: chenc68@zju.edu.cn;
zhaoxinkui@zju.edu.cn; guanjiech@126.com; dengsg@zju.edu.cn; zjuyjw@cs.zju.edu.cn).}

\thanks{Yueshen Xu is with the School of Computer Science and Technology, Xidian University, Xi’an 710126, China (e-mail: ysxu@xidian.edu.cn).}
}

%



\maketitle

\begin{abstract}
Interconnection is crucial for computing systems. However, the current interconnection performance between processors and devices, such as memory devices and accelerators, significantly lags behind their computing performance, severely limiting the overall performance. 
To address this challenge, Intel proposes Compute Express Link (CXL), an open industry-standard interconnection. With memory semantics, CXL offers low-latency, scalable, and coherent interconnection between processors and devices.
This paper introduces recent advances in CXL-based computing systems from single-machine to distributed. 
In single-machine systems, we classify existing research into two categories: Memory Expansion and Unified Memory. 
Memory Expansion focus on processors and memory, aims to address memory wall challenge.
Unified memory focus on processors and accelerators, aims to enhance collaboration in heterogeneous computing systems.
In distributed systems, we present how to build efficient disaggregation systems based on CXL infrastructure, enabling resource pooling and sharing. 
Finally, we discuss the future research and envision memory-centric computing with CXL.
\end{abstract}


\section{Introduction}
%
%
%
%

\IEEEPARstart{I}{nterconnection} is becoming a performance bottleneck in computing systems. A complex and heterogeneous computing system consists of Computing Units, such as CPU, GPU and TPU, Storage Units, Interconnections, such as PCIe and NVlink. 
The performance of computing units has grown 60,000 times in 20 years, yet the bandwidth of the interconnect has only grown 30 times during the same time\cite{memorywall}. Lagging interconnect performance results in the compute units needing to waste a lot of performance to reach synchronisation. 

In single-machine computing systems, interconnections can be categorized into two types: interconnections between processors and memory buffers, interconnections between processors and heterogeneous accelerators. 
The bottleneck between the processors and memory buffers result in \textbf{Memory Wall}, the pin-inefficiency of DDR memory limits the bandwidth of the processor to access the memory, as well as the memory capacity a processor can access\cite{intro_to_cxl}.
The bottleneck between processors and heterogeneous accelerators leads to a lack of efficient collaboration. The root cause is the \textbf{Non-Coherent Access} from device to device and device to processor. Data has to be moved to device-attached memory before computation, resulting in a lot of data movement.

With the rapid growth in the scale of data center applications (e.g., Deep Learning, Big Data), single-machine computing systems are unable to meet application`s demands in terms of computing power and storage capacity. Computing systems are gradually evolving towards distributed and large-scale.
In distributed computing systems, \textbf{Disaggregation}\cite{Disaggregation_survey} and \textbf{Sharing}\cite{DSM_survey} can effectively improve the efficiency and overall resource utilization. 
However, the current distributed interconnection is built on the basis of network (TCP, RDMA, etc.). Network`s high latency greatly hinders resource disaggregation and sharing. The overhead of a computing unit accessing computing and storage resources on other nodes through the network is much higher than accessing local resources.




To address these challenges, Intel proposed Compute Express Link (CXL)\cite{cxl1}, \cite{cxl2}, \cite{cxl3,cxlintro2}, an open industry-standard interconnection in 2019. CXL offers coherency and memory semantics with bandwidth that scales with PCIe bandwidth while achieving significantly lower latency than PCIe\cite{intro_to_cxl}.

CXL defines a set of protocols between processors and devices: CXL.io, CXL.mem, CXL.cache. 
The CXL.io protocol is based on PCIe. it is used for device discovery, status reporting, virtual-to-physical address translation, and direct memory access\cite{intro_to_cxl}.
CXL.cache is used by devices to cache system memory. 
CXL.mem enables processors to access the device-attached memory as cachable memory, called Host-managed Device Memory(HDM), thus enabling a unified view of the host between HDM and host memory.


\begin{figure*}[!t]
\centering
\includegraphics[width=0.99\textwidth]{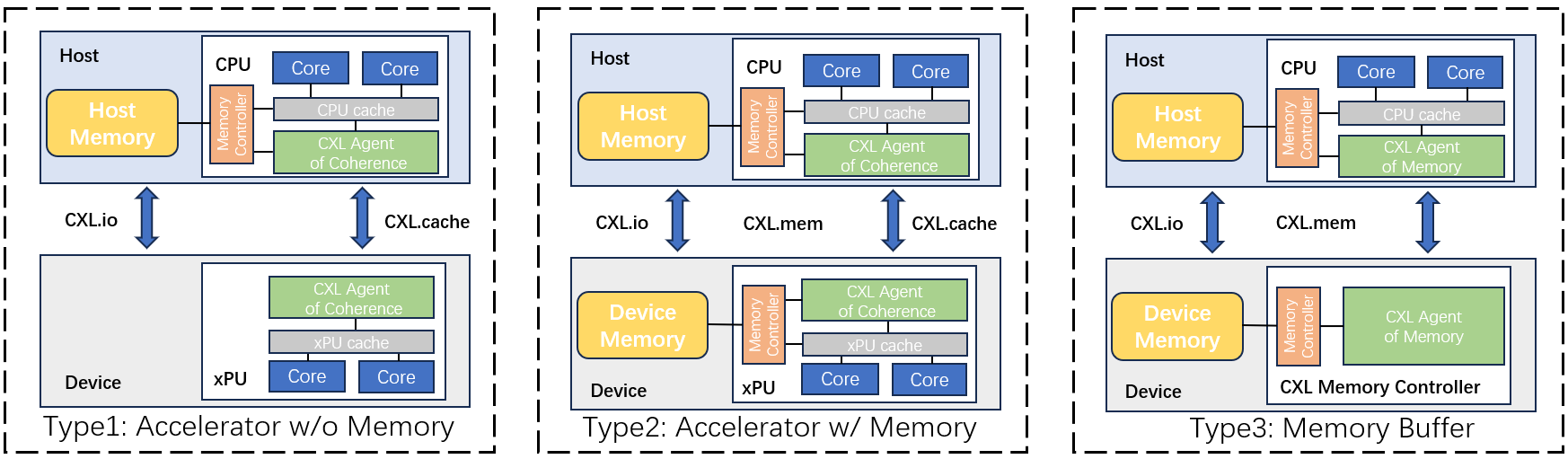}
\caption{Three types of CXL devices. The Type 1 device can directly cache host memory, the Type 2 device can directly cache each other's memory with the host, and the host can directly cache memory on a Type 3 device.}
\label{3typedevice}
\end{figure*}

Based on the combination of three protocols, CXL proposes three types of devices to address different scenarios, as shown in figure \ref{3typedevice}. 
Type 1 devices are accelerators without memory (e.g., NICs), which implement only the CXL.io and CXL.cache protocols, using coherency semantics along with PCIe-style DMA transfers\cite{intro_to_cxl}.
Type 2 devices are accelerators with memory (e.g., GPUs), which implement all CXL protocol and is able to cache each other's memory with the processor.
Type-3 devices implement only the CXL.io and CXL.mem protocols for memory bandwidth and capacity expansion. Compared to DDR memory, Type-3 devices are more cost, power and pin-efficient\cite{intro_to_cxl}.

Since 2019, CXL has evolved from 1.0 to 3.2. 
The 1.0 version adds cache coherence and memory semantics, solving the problem of fine-grained data sharing in heterogeneous computing, addressing the challenge of memory wall. 
The 2.0 version enables memory pooling with CXL Switch, creating a network of hosts and memory devices. 
The 3.0 version expands the memory pool by using multiple levels of CXL Switches, enabling the construction of dynamic, composable systems at the rack level. Additionally, The 3.0 version implements fine-grained memory sharing across host.

Prior to this survey, researchers from Microsoft published an introduction to CXL\cite{intro_to_cxl}. However, their work has two limitations. 
Firstly, the timing is too early. CXL Research has exploded in 2023 and 2024, only 10 paper were published during 2019 to 2022, while 40 in 2023, 51 in 2024. From another perspective, it means that CXL is being widely valued and shows great research potential.
Secondly, their work focus primarily on the CXL protocol itself, while barely touching on the research or real-world use cases. 

Our survey aims to fill the gaps left by previous survey, focusing on the latest CXL research and real-world use case, rather than delving deeply into the underlying hardware design.
We research the latest works (to the end of 2024) and present them with memory semantics, from single-machine to distributed system.

In single-machine computing system, we start with the interconnection between processors and memory buffers (Type3 Device), discussing how to use CXL memory for capacity expansion\cite{TPP, AccessLatencyisKey} and optimize workload`s performance\cite{CXL-ANNS,CXL-ANNS2}. 
Then, in order to avoid the overhead of data movement between processors and memory buffers. we discuss the \textbf{Near-Memory processing} \cite{NeoMem,recxl}, which focuses on how to offload computing functions to the memory side.


In addition to memory buffer, we also study the interconnection between processors and heterogeneous accelerators . CXL`s hardware cache coherence allows devices to uniformly cache data on each other’s memory, which not only avoid unnecessary data copying \cite{BreakingBarriers,Kwon2023FailureTT}, but also provides more memory capacity for heterogeneous devices\cite{GPUGraphProcessing}.

From single-machine to distributed system, CXL 2.0 proposes \textbf{Memory Pooling} with the CXL Switch. Our survey discusses how to decouple compute and memory resources from nodes and build a fully disaggregation system \cite{pond},\cite{DirectCXL,ddc}, enabling flexible and efficient resource utilization.
We also focus on CXL based \textbf{Distributed Shared Memory}. In CXL 3.0, a memory module can be shared by multiple compute modules, which inspires researchers to use shared memory for distributed communication \cite{NotNets,cxlshm} and optimizing the performance of in-memory workloads such as redis\cite{redis}.

Our survey provides a comprehensive review of existing CXL systems, classifies and summarizes these studies based on interconnection type, research contents and workloads. 
Then, we discuss the limitations of the existing works and the possible directions for future research. 
Finally, we envision a memory-centric computing paradigm based on CXL fabrics.

The rest of the survey is organized as follows. 
Section \ref{section:Evaluation and Simulation} introduces the performance evaluation of existing CXL systems, the real world use cases, and the simulation platform for CXL hardware.
Section \ref{section:Memory Expansion} introduces the memory expansion with CXL Type3 Device.
Section \ref{NMP} discusses how to avoid the costly data movement with near-memory computing.
Section \ref{section:Unified Memory} presents a discussion about how to use CXL to accelerate heterogeneous computing systems. 
Section \ref{section:Memory pooling} talks about CXL based disaggregated system.
Section \ref{section:Distributed Shared Memory} introduces the researches on distributed shared CXL memory.
In Section \ref{Future Research}, we propose five main directions for future research.
In Section \ref{Memory-Centric Computing}, we introduce the memory-centric computing paradigm in the CXL era. 
Finally, Section \ref{conclusion} gives a conclusion.
Table \ref{tab: overview} shows an overview of CXL based computing systems from single-machine to distributed.

\begin{table*}[htbp]
    \centering
    \caption{An Overview of CXL Based Computing Systems}
        \begin{tabular*}{0.95\linewidth}{cccc}
            \toprule
            \textbf{Scale} & \textbf{Classification} & \textbf{Research Topics} & \textbf{Reference} \\
            \midrule
            \multirow{4}{*}{Single-machine and Distributed} & \multirow{4}{*}{Evaluation and Simulation} & The Latency and Bandwidth. &\cite{TPP},\cite{pond},\cite{sun2023demystifying},\cite{tang2024exploring},\cite{liu2024exploring} \\
            & & Use Case: Regular Applications. &\cite{wahlgren2022evaluating},\cite{sun2023demystifying},\cite{tang2024exploring},\cite{liu2024exploring}  \\
            & & Use Case: Large Language Models. 
            & \cite{tang2024exploring},\cite{liu2024exploring},\cite{ZeRO-Offload},\cite{FlexGen} \\
            & & The Simulation Platforms &\cite{wahlgren2022evaluating},\cite{cxlHybridPool},\cite{yang2023cxlmemsimpuresoftwaresimulated},\cite{gond2024emucxlemulationframeworkcxlbased},\cite{mess},\cite{DRackSim}  \\
            \midrule
            \multirow{7}{*}{Single-machine} & \multirow{3}{*}{Memory Expansion} & Tiered Memory System
            & \cite{TPP}, \cite{AccessLatencyisKey},  \cite{Nomad,SMT},  \cite{song2023lightweightfrequencybasedtieringcxl},  \cite{tirumalasetty2024exploringdramcacheprefetching} \\
            & & Flash Memory as a Backend & \cite{cxlssd,CacheinHand},\cite{Hellobytes} \\
            & & Application-Specific Optimizations 
            & \cite{CXL-ANNS,CXL-ANNS2}, \cite{DeepMemoryDL}, \cite{recxl}, \cite{tang2024exploring,CXLKVLLM} \\ \cmidrule{2-4}
            & \multirow{2}{*}{Near-Memory Processing} & Workload-Customized Near-Memory Process 
            & \cite{clay},\cite{BEACON},\cite{CXL-PNM},\cite{recxl} \\ 
            & & Generalized Near-Memory Process 
            & \cite{cms}, \cite{Polaris},\cite{NeoMem},\cite{M2NDP},\cite{UDON} \\ \cmidrule{2-4}
            & \multirow{2}{*}{Unified Memory} & Memory Expansion with CPU Relay 
            & \cite{GPUGraphProcessing},\cite{LMB} \\
            & & Unified Memory with Direct CXL Access &\cite{DesignandanalysisCXL},\cite{SynergizingCXLwithUnified},\cite{BreakingBarriers},\cite{Kwon2023FailureTT},\cite{Arif2023AcceleratingPO},\cite{salus} \\
            \midrule
            \multirow{6}{*}{Distributed} & \multirow{3}{*}{Memory Pooling} & The Architecture of Pooling System 
            & \cite{DirectCXL},\cite{MemoryPoolingWithCXL},\cite{ddc},\cite{Amaro2023LogicalMP},\cite{FFC} \\
            & & The Interconnection in Pooling System &\cite{Geyer2023WorkingWD},\cite{NeartoFar},\cite{Aurelia},\cite{CXLoverEthernet},\cite{rcmp},\cite{rPCIeBench} \\ 
            & & The Cost of Pooling System 
            & \cite{ACaseAgainstCXL},\cite{pond,pond2}\\ \cmidrule{2-4}
            & \multirow{3}{*}{Memory Sharing} & Distributed Shared Memory with CXL 3.0 
            & \cite{jain2024memorysharingcxlhardware},\cite{cxlshm},\cite{TrEnv} \\
            & & Communication via Shared CXL Memory 
            & \cite{cxlshm},\cite{HydraRPC},\cite{NotNets},\cite{rpcool}\\ 
            & & Failure Tolerance & \cite{cxlshm},\cite{xu2024cxlsharedmemoryprogramming} \\ 
            \bottomrule
        \end{tabular*}
    \label{tab: overview}
\end{table*}




\section{Evaluation and Simulation}
\label{section:Evaluation and Simulation}

Before discussing the existing research, we first introduce the evaluation results of real CXL systems. 
Existing work \cite{liu2024dissectingcxlmemoryperformance,wang2024hitchhikersguideprogrammingoptimizing}, \cite{sun2023demystifying,tang2024exploring}, \cite{liu2024exploring} evaluates CXL from two perspectives: the basic performance and real-world use case.
Based on the results, researchers not only can gain a deeper understanding of the performance characteristics of CXL hardware, but also can guide the construction of their own experimental environment.
This section mainly introduces the results of \cite{sun2023demystifying,tang2024exploring}, \cite{liu2024exploring}.

\subsection{The Latency and Bandwidth of Real CXL Hardware.} 
In 2023, researchers from the University of Michigan propose TPP\cite{TPP}, a page placement approach for the tiered memory system which includes CXL memory. TPP provides the latency characteristics of CXL memory and other memory technologies, which have been widely referenced, as shown in the following table \ref{Characteristics of Memory}.

\begin{table}[htbp]
    \centering
    \caption{Characteristics of Memory Technologies.}
    \label{tab:advanced_example}
        \begin{tabular*}{0.8\linewidth}{c|c|c}
            \toprule
            \textbf{Memory Technologies  } & \textbf{Capacity} & \textbf{Latency}\\
            \midrule
                Register & B & 0.2ns\\
                CPU Cache & KB-MB & 1-40ns\\
                Main Memory & GB-TB & 80-140ns\\
                CXL Memory & TB & 170-250ns\\
                NVM & TB & 300-400ns\\
                Far Memory & GB-TB & 2-4us\\
                SSD & TB & 10-40us\\
                HDD & TB-PB & 3-10ms\\
            \bottomrule
        \end{tabular*}
        \label{Characteristics of Memory}
\end{table}


In the same year, Pond\cite{pond} from Microsoft Azure not only measures the end-to-end latency of host direct access to CXL memory, but also states the tradeoff between latency and scale. In Pond's evaluation, in an 8-socket system, the latency of the host direct access to the CXL memory controller is \textbf{155ns}, while in a 16-socket system, the latency of access to a more distant CXL memory controller through a retimer is \textbf{180ns}. In larger systems (32-64 sockets), the latency of connecting through retimer + CXL switch will exceed \textbf{270ns}.

The researchers from UIU and Intel\cite{sun2023demystifying} evaluate and compared the performance of real CXL hardware from three different vendors. Each device has different CXL IP (ASIC-based hard IP and FPGA-based soft IP) and DRAM technologies (DDR5-4800, DDR4-2400, and DDR4-3200). This work proposes several important conclusions:
\begin{itemize}
\item[(1)] Real CXL hardware can provide 26\% lower latency and 3-66\% higher bandwidth than simulated CXL memory, as real CXL hardware does not have caches or CPUs, avoiding the overhead of maintaining cache coherency.
\item[(2)] While real CXL hardware is exposed as a NUMA node, it does not require LLC isolation, which means CXL memory may have several times the LLC capacity of regular memory.
\end{itemize}

The researchers from ByteDance and Harvard\cite{tang2024exploring} focus on the currently commercially available ASIC-based CXL Type3 devices, and conduct a more in-depth evaluation, proposing several new conclusions.

The micro-benchmark results show that the ASIC-based CXL Type3 device significantly outperforms the FPGA-based CXL Type3 device. Interestingly, the researchers find that the location of the CXL hardware is very important: When the CXL device is located on the local socket, its latency is around \textbf{256ns}, which increases slightly with the ratio of read\&write operations. When the CXL hardware is on a remote socket, the latency increases sharply to \textbf{485ns}. Additionally, as the access bandwidth increases, the latencies of both CXL and DRAM gradually increase, and when reaching a certain threshold, the latency will increase significantly.

Jie Liu et al\cite{liu2024exploring} also evaluate CXL hardware and make more in-depth conclusion. The result tells that the bandwidth of CXL hardware is much lower than regular memory, limited by PCIe (the test platform is PCIe Gen5), with CXL hardware able to sustain around 8 threads (regular memory is 28 threads), and the peak bandwidth of CXL hardware is only about 40\% of regular memory.

\subsection{The Real-world Use Case: Regular Applications.}
Jacob Wahlgren et al.\cite{wahlgren2022evaluating}provide a CXL simulation platform and an analysis tool to identify the memory usage patterns of applications and their optimization opportunities. Their evaluation uses two modes: 
\begin{itemize}
\item[(1)] The first mode keeps the total bandwidth consistent but changes the ratio of DRAM and CXL memory, using scientific computing applications\cite{Asanović:EECS-2006-183} as the workload.
\item[(2)] The second mode maintains a consistent memory ratio but varies the CXL memory bandwidth, using graphics applications as the workload.
\end{itemize}
The evaluation results show that even with 75\% of the data placed in the pooled memory, the performance impact on some applications is less than 10\%. Furthermore, the study finds that dynamically configuring high-bandwidth systems by increasing the number of CXL links can effectively support bandwidth-intensive applications, such as OpenFOAM based on unstructured grids.

Researchers from UIU and Intel\cite{sun2023demystifying} evaluate real cxl hardware on latency-sensitive applications(Redis\cite{redis} with YCSB\cite{YCSD}, FIO\cite{fio}, etc.) and throughput applications(DLRM inference\cite{MERCI}, etc.). The results show a 10-82\% performance degradation when running applications directly on CXL memory. Even with dynamic page migration techniques like TPP\cite{TPP}, it can still increase tail latency.

Researchers from ByteDance and Harvard\cite{tang2024exploring} conduct detailed evaluations on four real-world workloads: KVS, Spark, ECS, and CPU-based LLM inference. The consistent conclusion is that the mixed use of CXL memory and DRAM(Memory Interleave) does cause some performance degradation, but it is better than using SSDs. Tiered memory strategies like TPP\cite{TPP} can effectively improve application performance in the KVS scenario, but perform poorly in data analysis scenarios like Spark.

Jie Liu et al.\cite{liu2024exploring} also evaluate CXL hardware on HPC applications. Similar to previous work, compute-intensive applications with low memory access can benefit the most from CXL, with relatively small performance degradation. Bandwidth-intensive applications like data analytics, however, will be greatly impacted.

\subsection{The Real-world Use Case: Large Language Models.}
The researchers from ByteDance and Harvard\cite{tang2024exploring} design a evaluation of CPU-based LLM inference workload, which leads to an important conclusion: for compute-intensive applications, the mixed use of CXL memory and DRAM can outperform both \textbf{DRAM Only} and \textbf{Tiered Memory}. The key reason is that the bandwidth of DRAM is limited, and when the access density reaches DRAM`s threshold, the efficiency drops sharply. By offloading part of the memory access load to CXL memory, avoiding the DRAM`s performance degradation, the overall application performance can be improved. This has led researchers to rethink whether the Tiered Memory (using DRAM as a cache for CXL memory) needs to be upgraded.

Jie Liu et al\cite{liu2024exploring}. also discuss the impact of CXL on large language models (LLMs) in depth, from three perspectives. First, tensor offloading, where objects are offloaded from GPU memory to CXL memory, but due to the lack of CXL-enabled accelerators, this has to go through a relatively long path (GPU-CPU-CXL), severely impacting performance. Second, based on ZeRO-Offload\cite{ZeRO-Offload}, they evaluate the impact of CXL on the LLM training, concluding that compared to regular memory, CXL causes performance degradation in training, as parameter computation requires frequent access to the slower CXL memory. Finally, based on FlexGen\cite{FlexGen}, they evaluate the impact of CXL on the LLM inference, finding that the Decode stage is well-suited for CXL as it revolves around the KV cache without intensive memory accesses, while the Prefill stage sees performance degradation due to both latency and bandwidth limitations of CXL.

\subsection{The Simulation Platform}
Since most of the CXL hardware are currently in the prototype stage, only a few products have been commercialized. Therefore, it is not easy for researchers to acquire CXL hardware. Fortunately, some simulation platforms targeting CXL hardware has been proposed, with their help, researchers can simulate a system which equipped with CXL hardware. In this section, we will introduce these simulation platforms.

Jacob Wahlgren et al.\cite{wahlgren2022evaluating} implement a cxl memory simulation tool based on the libnuma library, using remote numa node to simulate CXL memory. 

Researchers from Samsung, Qirui Yang et al.\cite{cxlHybridPool}, also based on the NUMA architecture, use remote NUMA node to simulate CXL hardware. Unlike other work, Yang attempts to build a tiered memory architecture to simulate not only DRAM memory but also NVMe SSD on the simulated CXL NUMA node. Applications or VMs prioritize accessing memory on CXL devices, and when CXL memory is exhausted, they will use the Linux kernel's Swap mechanism to swap memory pages between CXL memory and SSD.

Yiwei Yang from UCSC proposes \textbf{CXLMemSim}\cite{yang2023cxlmemsimpuresoftwaresimulated}, a pure software-simulated tool for analyzing the performance characteristics of CXL memory. CXLMemSim not only can be used to simulate CXL memory, but also proposes an event-driven performance model that attaches to existing unmodified programs, dividing program execution into multiple phases and collecting performance monitoring events at the end of each phase. Based on these events, it calculates the execution time to evaluate the performance impact of CXL memory on the application.

Researchers from IIT, Raja Gond et al., propose  \textbf{emucxl}\cite{gond2024emucxlemulationframeworkcxlbased}, an emulation framework for CXL memory. emucxl provides a user-space library and a NUMA-based CXL simulation backend, allows developers and researchers to rapidly prototype, without the need to rebuild the CXL platform and access abstraction layers. The work also demonstrates the application of emucxl in two different use cases: direct application access and application middleware.

Pouya et al. propose the  \textbf{Memory Stress} (Mess) framework\cite{mess}, which provides a comprehensive evaluation of almost all memory systems, including the simulation and analysis of the CXL memory. They simulate the CXL memory based on the bandwidth-latency curves provided by the manufacturers. 
They have made the Mess framework's source code publicly available, providing the community with a powerful tool for researching and utilizing these new memory technologies.

Researchers from IIT, Amit Puri et al., propose \textbf{DRackSim}\cite{DRackSim}, a simulation infrastructure for modeling large-scale disaggregated memory systems supporting CXL. 
DRackSim simulates multiple compute nodes, memory pools, local/global memory managers, and interconnects that support coherent memory access. 
Additionally, DRackSim integrates a modified version of \textbf{DRAMSim2} to enable simulation of local and remote memory.
DRackSim provides a powerful tool for researchers to evaluate and optimize new designs for large-scale disaggregated memory systems in data centers.

\section{Memory Expansion}
\label{section:Memory Expansion}

\begin{figure}[!t]
\centering
\includegraphics[width=0.49\textwidth]{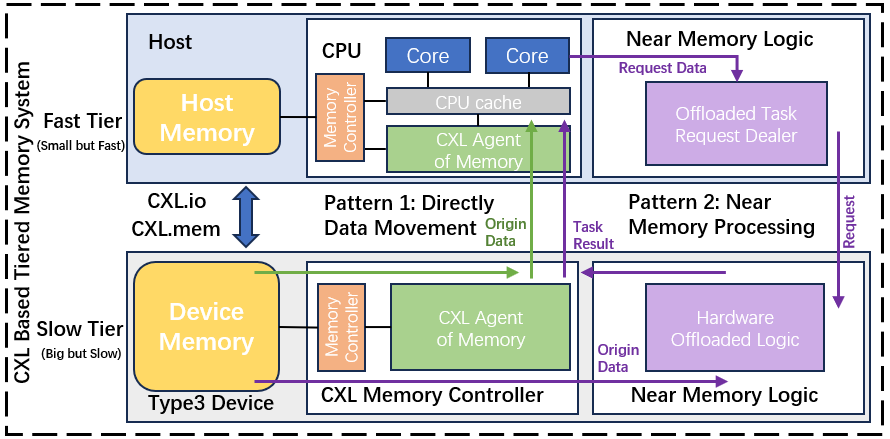}
\caption{CXL-based Memory Expansion. It is based on tiered memory, which places data in appropriate memory tiers. In order to avoid the overhead of data movement between tiers, researches introduce near-memory processing.}
\label{memory_expansion}
\end{figure}

As shown in figure \ref{memory_expansion}, the most common application scenarios is using CXL Type3 Device for \textbf{Memory Expansion}, which breaks the physical constraints of a single machine, allows hosts to access the huge memory on the device through Load/Store semantics and with lower latency compared with other memory expansion approaches(NVMe, RDMA, etc.). Currently, there is a lot of work on using CXL to alleviate the memory bottleneck faced by memory intensive applications, such as ANN, Large Recommendation Systems, Graph Processing, and In-Memory Databases.


\subsection{Tiered Memory System}
Although CXL Memory has a larger capacity, however, its latency is much higher than DRAM, which causes performance degradation.
Tiered Memory is an OS-level mechanism for leveraging heterogeneous memory, especially CXL-enabled systems. 
The current tiered CXL Memory systems\cite{TPP}, \cite{AccessLatencyisKey},  \cite{Nomad,SMT},  \cite{song2023lightweightfrequencybasedtieringcxl},  \cite{tirumalasetty2024exploringdramcacheprefetching}, represented by the TPP\cite{TPP}, efficiently places hot/cold pages in appropriate memory tiers, ensuring performance-critical pages are in fast DRAM memory while offloading less critical ones to slower CXL Memory. It operates transparently without application-specific knowledge and has been integrated into the Linux kernel.

Lingfeng Xiang et al. find out that \textbf{exclusive memory tiering}, in which a page is either present in fast memory or slow memory, but not both simultaneously, is not the best strategy for tiered memory management. 
Thet propose \textbf{Nomad} \cite{Nomad}, a new page management mechanism for Linux that features transactional page migration and page shadowing. NOMAD helps remove page migration off the critical path of program execution and makes migration completely asynchronous. 

Midhul Vuppalapati et al. argue that the basic assumption that tiered memory systems rely on: \textbf{DRAM latency is always lower than CXL memory}, needs to be reconsidered. 
They find that DRAM`s latency grows rapidly once the DRAM is highly loaded, e.g. when the bandwidth is close to saturation. 
They proposed Colloid\cite{AccessLatencyisKey},  a memory management mechanism that embodies the principle of balancing access latencies: page placement across tiers should be performed so as to balance their average (loaded) access latencies.

Researchers from Microsoft Azure propose CXLinCloud series \cite{pond,pond2}, \cite{CXLINVIRTUAL}. 
This series focuses on introducing CXL pooling memory in \textbf{cloud computing}.
They first propose Pond \cite{pond,pond2}, a CXL memory pool for cloud. Pond meets performance goals by using machine learning models trained on historical data (cloud production traces) to guide page migration and memory allocation in a tiered CXL memory systems. Pond predicts the amount of local and pool memory needed for virtual machines, resembling same-NUMA-node memory performance. 
However, in virtualized environments, software-based tiering introduces high overheads, hardware-based tiering can not deal with multi-tenant.
They introduce Memstrata \cite{CXLINVIRTUAL}, a lightweight multi-tenant memory allocator. Memstrata employs page coloring to eliminate inter-VM contention. It improves performance for VMs with access patterns that are sensitive to hardware tiering by allocating them more local DRAM using an online slowdown estimator.

\subsection{Flash Memory as a Backend}
CXL doesn't limit storage technology, and researchers are looking at ways to use flash memory as main memory through CXL, which are more cost-effective than expensive DRAMs, to store data.

Works represented by the CXL-SSD\cite{cxlssd,CacheinHand},\cite{Hellobytes} poses three major challenges of using flash memory: 
\textbf{granularity mismatch between CPU and SSD}, which leading to traffic and write amplification, 
\textbf{significant latency compared to DRAM}, 
and \textbf{limited endurance} due to a finite number of program/erase cycles. 

To mitigate these challenges, researchers explore caching strategies, including the use of a DRAM cache and MSHRs to reduce repeated reads, and prefetching techniques to hide flash memory latency. They also evaluate different flash memory technologies and parallelism levels, finding that ULL and SLC perform better in terms of latency and endurance. Lastly, they suggests system-level changes, such as passing memory access hints from the kernel to the CXL-SSD, to improve prefetching accuracy and achieve performance closer to DRAM.

\subsection{Application-Specific Optimizations}
Tiered CXL memory systems require specific memory policies for different application to maximize performance. In this section, we present CXL systems designed for four major application scenarios, namely Approximate Nearest Neighbor Search(ANNS), Deep Learning, Database(DB), Serverless Computing.

Works represented by CXL-ANNS\cite{CXL-ANNS,CXL-ANNS2,cheng2024characterizingdilemmaperformanceindex} using CXL extended memory to optimize the efficiency of large-scale ANN search. Firstly, they put complete datasets required for ANNS into CXL memory; 
then, they focus on tuning tiered CXL memory with cache and prefetching. Prefetching policies are set up to address the localized nature of ANNS, and caching policies are set up to address the relationship between ANNS nodes; 
lastly, they address the fact that the CXL memory pool consists of multiple CXL devices (the complete dataset can be distributed on multiple CXL devices), the process of ANNS is optimized for parallel computation to further improve the performances.

Deep learning(DL) workloads also have significant memory requirements, especially \textbf{Large Recommendation System} and \textbf{Large Language Model}.
Moiz Arif et al. are the first to use CXL for DL workload`s memory extension. They propose DeepMemoryDL \cite{DeepMemoryDL}, which manages the allocation of additional CXL-based memory and provides intelligent prefetching and caching mechanisms for DL workloads.
Dong Xu et al.\cite{TensorOffloading} improve the existing tensor-offloading approach. They use CXL to build a unified memory between main memory and accelerator memory, enabling accelerators to use main memory directly, addressing the storage challenge of deep learning workloads
Miryeong Kwon et al. propose \textbf{TRAININGCXL} \cite{Kwon2023FailureTT} that can efficiently process large-scale recommendation datasets in CXL memory pool while making training fault tolerant with low overhead.
Haifeng Liu et al. introduce ReCXL\cite{recxl}, which utilizes near-memory processing for scalable, efficient recommendation model training. 
Yupeng Tang et al. explore leveraging CXL memory to store KV cache in LLM serving \cite{tang2024exploring,CXLKVLLM}.
Since DL workloads often rely on heterogeneous computing systems, we will detail related works in section \ref{NMP} and section \ref{section:Unified Memory}.

In database, works\cite{cxldb01},  \cite{cxldb02,cxldb03},\cite{Lee2024DatabaseKS} using a probabilistic page migration policy to optimize data locality, and exposes memory tiers as NUMA nodes for performance tuning. 
Demand paging allocates pages to local or CXL memory, managing memory with a single copy to simplify parallel access. 
Variable page sizes are supported to prevent external fragmentation, and state management is achieved through frame metadata storage. 
The eviction strategy employs a second-chance FIFO queue to mimic LRU, with pages evicted to SSD using specific system calls. 
These designs effectively use tiered CXL memory to optimize the database system, enhancing scalability and efficiency across memory tiers.

Works represented by Apta\cite{cxlfass02}, \cite{TrEnv},  \cite{cxlfass03,li2023understandingoptimizingserverlessworkloads} using CXL to enhance the performance of serverless applications by addressing the limitations of remote object stores and the lack of fault tolerance. They employ a simplified two-state coherence protocol, allowing compute servers to cache objects without blocking in the event of server failures. Additionally, They optimize data movement by supporting object-granular reads and writes, which are more efficient for FaaS object access patterns compared to CXL's cache-line granular accesses, through bulk cache line loading and transactional atomic durability, reducing the overhead of software transactions.

\section{Near-Memory Processing}
\label{NMP}

The reason for the high latency of accessing CXL memory is that CXL Type3 Device itself does not have computing capability, the data must go through a costly movement from device to processor to be processed.
A possible solution is \textbf{Near-Memory Processing}(NMP). Pioneer researchers are working to introduce NMP to CXL systems, offloading some computing functions to the CXL memory devices, so that the computation can be executed on the memory side (as shown in figure \ref{memory_expansion}), avoiding data movement.


\subsection{Workload-Customized Near-Memory Process}
Sungmin Yun et al. propose \textbf{CLAY}\cite{clay}, a scalable NDP architecture based on CXL, aiming to accelerate the key building blocks of deep neural networks (DNNs) - the embedding layer. 
CLAY reduces the data transfer overhead by interconnecting DRAM modules, and designs a specialized memory address mapping to alleviate the load imbalance problem. Additionally, CLAY introduces a data packet replication scheme to reduce the demand for instruction transfer bandwidth.

Wenqin Huangfu et al. propose \textbf{BEACON} \cite{BEACON}, scalable Accelerators for Genome Analysis with near CXL memory processing. 
Exsiting systems have two critical limitations, i.e., performance bottle-necked by communication and the limited potential for memory expansion. 
They first implement a CXL2.0 prototype to expand memory capacity, where the host connects to multiple CXL DIMMs through a CXL Switch. The NDP module is ported to the CXL Switch. A data allocation and migration method has been designed to minimize communication (data movement, etc.) between DIMMs, thereby avoiding the communication problems.



Nam Sung Kim et al. in collaboration with Samsung, propose \textbf{CXL-PNM}\cite{CXL-PNM}, a near-data computing platform based on LPDDR, aimed at providing high-performance inference acceleration for Transformer-based large language models (LLMs). CXL-PNM utilizes LPDDR5X DRAM to provide up to 512GB of capacity and 1.1TB/s of bandwidth. 
The platform features a CXL controller architecture that integrates an LLM inference accelerator. This controller not only implements the CXL protocol and integrates the PNM accelerator, but also employs a hardware arbiter to efficiently manage concurrent memory access requests from the CPU and PNM accelerator. 

Haifeng Liu et al. propose \textbf{ReCXL} \cite{recxl}, a novel architecture aims at improving the efficiency of large-scale recommendation model training. ReCXL addresses the performance bottlenecks by leveraging CXL and NDP architectures. The NDP architecture of ReCXL processes the entire embedding training process within the CXL memory, and combines software-hardware co-optimization techniques, such as dependency-free prefetching and fine-grained update scheduling. 

\subsection{Generalized Near-Memory Process}
Joonseop Sim et al. from SK Hynix propose a solution called \textbf{CMS}\cite{cms} to address the bandwidth limitation of CXL devices. CMS integrates NDP cores in the CXL device, which reduces the data movement through the CXL interface. CMS also proposes methods to improve CMS bandwidth utilization through hardware logic load balancers and MAC block optimizations.

Zhe Zhou et al, propose \textbf{Polaris}\cite{Polaris}, which integrates a hardware prefetcher into the CXL memory controller to reduce the latency of accessing CXL memory, without requiring modifications to the CPU or software. Polaris analyzes memory requests and prefetches cache lines into a dedicated SRAM buffer. 
The dedicated buffer avoids polluting the CPU cache.

Based on Polaris, they propose \textbf{NeoMem}\cite{NeoMem}, a hardware-software co-designed CXL-based tiered memory system. NeoMem aims to enhance the efficiency of existing memory tiering systems by integrating a dedicated hardware unit called NeoProf in the CXL controller. 

NeoProf is capable of real-time tracking of memory accesses, providing the operating system with critical page hotness statistics and other useful system state information. Zhou implements a dynamic hot page promotion strategy in the Linux kernel, based on NeoProf's statistical data. 
Considering the dynamic characteristics of memory access patterns, the identification threshold for hot pages is adjusted in real-time. 
Additionally, NeoMem monitors the bandwidth usage of the slow CXL memory, and when high bandwidth utilization is detected, it triggers more pages to be migrated to the fast memory tier. 
To avoid frequent page migrations, NeoProf introduces a page demotion flag to measure the severity of the thrashing phenomenon, and adjusts the threshold accordingly.

Hyungkyu Ham et al. from POSTECH propose \textbf{M2NDP} (Memory-Mapped NDP) \cite{M2NDP}, designed specifically for CXL memory expanders. M2NDP achieves low-cost, general-purpose NDP functionality by combining memory-mapped functions (M2func) and memory-mapped micro-threading (M2thr). M2func utilizes the unmodified CXL.mem protocol to enable lightweight communication between the host and CXL devices, avoiding the high overhead of traditional PCIe/CXL.io solutions. M2thr introduces micro-threads, a lightweight thread with minimal register allocation, allowing a large number of micro-threads to execute concurrently on low-cost NDP units. 

Jon Hermes et al. propose \textbf{UDON}\cite{UDON}, exploring the potential and benefits of offloading tasks to general-purpose cores on CXL memory devices. They simulate a CXL type-2 device (a device with general-purpose compute cores) on an Arm AArch64 dual-socket NUMA system. This study finds that by intelligently partitioning the machine learning inference model for computation offloading, up to 90\% of the data can be placed in CXL memory with only a 20\% performance loss.



\section{Unified Memory}
\label{section:Unified Memory}

\begin{figure}[!t]
\centering
\includegraphics[width=0.49\textwidth]{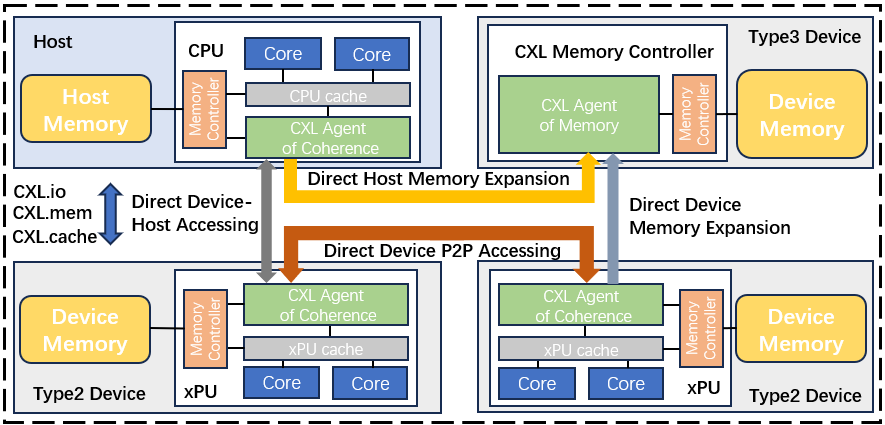}
\caption{Unified Memory. The hardware cache coherence provided by CXL organises the device-attached memory and host-attached memory into a unified memory space.}
\label{unified memory}
\end{figure}

CXL system include not only processors and memory devices, but also heterogeneous accelerators (Type2 Device, such as GPUs, TPUs, etc.). 
Unlike memory devices, accelerators have both compute units and memory.
However, due to the lack of cache coherence between accelerators and processors, the main memory and device memory cannot be accessed uniformly. 
The data interaction between devices requires the CPU to act as a data transfer relay, which has become a performance bottleneck in heterogeneous computing systems. 

As shown in figure \ref{unified memory}, CXL provides hardware-based cache coherence to enable the \textbf{Unified Memory} across accelerators and processors. 
It allows accelerators to uniformly access data on each other's memories, which not only avoid unnecessary data copying, improve the computing efficiency of the CPU, but also extend memory capacity for heterogeneous devices.


\subsection{Memory Expansion with CPU Relay}
In this section we focus on how to use CXL to expand the memory of the accelerators with the help of the CPU, without any hardware modifications.

Shintaro Sano et al. from Kioxia Corporation use CXL memory to address the memory requirements of GPU-based graph processing workloads\cite{GPUGraphProcessing}. This work uses commercial GPUs and FPGA-based CXL Memory (based on the Intel Agilex®7 FPGA) to build a heterogeneous memory system. 
Since commercial GPUs do not support the CXL standard, GPU`s memory requests have to be translated by the CPU to access the CXL memory, resulting in a longer access path. 
They also discuss the importance of small address alignment sizes and appropriate data transfer sizes for runtime optimization, and point out that in this GPU-CPU-CXL access pattern, the PCIe interface bandwidth to the GPU is the main performance bottleneck.

Jiapin Wang et al. propose \textbf{LMB}\cite{LMB}, aiming to address the DRAM capacity issues faced by SSDs and GPUs. LMB utilizes the low-latency feature of CXL to dynamically extend the memory of PCIe devices, allowing efficient host-forwarded sharing of memory resources between CXL memory and PCIe devices. They also discuss the challenges, such as dynamic memory allocation, resource optimization, shared resource isolation, access control, cross-device migration, and data security, and propose a unified memory allocation interface that allows PCIe and CXL devices to easily utilize LMB's memory resources, just like using on-board memory.

\subsection{Unified Memory with Direct CXL Access}
In this section we focus on how to using CXL to build a unified memory across heterogeneous accelerators, which requires GPU hardware support for the CXL protocol, known as a CXL-Type2 Device. 

In 2022, Anthony M Cabrera et al. design and analyze the performance model of CXL devices in heterogeneous computing systems with existing hardware \cite{DesignandanalysisCXL}. 
Since CXL device prototypes are not yet available, they develop a benchmark application called DecEval to simulate the collaborative work of GPUs and FPGAs using existing software stacks and frameworks. 
The CXL performance model is based on high-level expectations of how CXL would operate, including a theoretical analysis that compares the performance of CXL and PCIe. 
They use existing PCIe models as a comparison baseline and evaluate the CXL model based on measurement data collected from the benchmark application. 

The work by Junseung Lee et al.\cite{SynergizingCXLwithUnified} discusses how to expand the memory capacity of GPUs in a heterogeneous computing environment with CXL. 
They believe that providing a unified memory space abstraction can relieve programmers of the burden of manually managing different memory spaces and data transfers, with the system responsible for deciding the migration of data across different memories at the underlying level. 
They also analyze different memory configuration schemes and proposes the design space for a unified memory system, such as the location of data and how to migrate data between memory tiers.

The team led by Myoungsoog Jung delves deeper into how to use CXL to extend and share memory for heterogeneous accelerators in a unified memory manner\cite{BreakingBarriers}. They integrate CXL into the GPU. The GPU contains multiple CXL root ports, each equipped with a CXL controller that support DRAM or SSD. 
Through speculative reads and deterministic write mechanisms, allowing the GPU to predict and prefetch data, while also handling the latency of write operations. 
The GPU's system bus and memory mapping allow the GPU to directly issue memory requests, which are handled by the CXL root port, without the need for CPU involvement. 

Based on the previous work, the team from KAIST proposes \textbf{TRAININGCXL} \cite{Kwon2023FailureTT}, which uses CXL to integrate persistent memory and GPUs into a cache-coherent domain as a Type-2 device, allowing GPUs to directly access persistent memory without software intervention, enabling efficient large-scale recommendation model training. The system introduces computation and checkpoint logic near the CXL controller to proactively manage training data and persistence, optimizing the checkpoint overhead on the critical path of training. 

Moiz Arif et al. discuss the resource contention issues that arise when accelerating GPU-based workloads through CXL\cite{Arif2023AcceleratingPO}. They believe that the contention mainly occurs in two aspects.
First, the PCIe bandwidth: when multiple GPUs access CXL memory simultaneously, it can saturate the PCIe bandwidth.
Second, the memory utilization: traditional memory allocation and task scheduling can lead to imbalanced memory utilization among multiple GPUs
They propose a memory allocation strategy that considers the memory requirements of each job and the available memory resources. Based on the current memory usage of the jobs and the overall memory demand of the system, the strategy can also dynamically adjust the memory allocation to reduce memory contention. The strategy also considers the bandwidth limitations and access latency of the PCIe and CXL to avoid bandwidth contention. 

Dong Xu et al.\cite{TensorOffloading} improve the existing tensor-offloading approach. They use CXL to build a cache coherence domain between CPU memory and accelerator memory, slightly extend CXL to support an update cache-coherence protocol and avoiding unnecessary data transfers.

Rahaf Abdullah et al. propose a new security model called \textbf{Salus}\cite{salus} for unified memory. Unified memory requires dynamic migration of data across multiple heterogeneous storage, but in cloud GPU scenarios, existing security implementations are difficult to adapt to data migration, as their metadata is closely tied to the physical location of the data. Salus decouples secure computation from physical data location and designs structured encryption counter blocks to reduce traffic during data migration, and significantly reduces traffic related to metadata access and writeback by using a bitmask format for dirty information tracking in the CXL-to-GPU mapping. 
\section{CXL Based Disaggregated System}
\label{section:Memory pooling}

\begin{figure}[!t]
\centering
\includegraphics[width=0.49\textwidth]{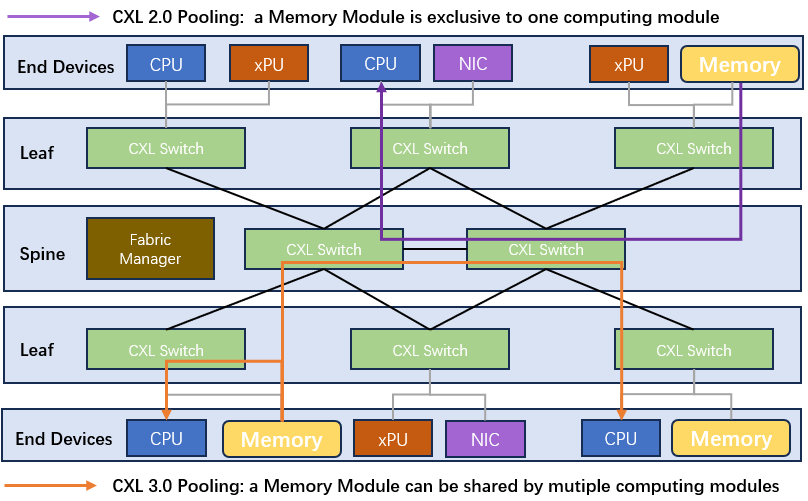}
\caption{CXL Based Disaggregated System\cite{intro_to_cxl}. It breaks down the physical isolation between servers and consolidates the data center into a hyper-node through resource pooling.}
\label{pooling memory}
\end{figure}

Based on the memory expansion on single-machine, CXL 2.0 proposes a solution for \textbf{Memory Pooling} with the CXL Switch, as shown in figure \ref{pooling memory}. 

In data centers, servers will be decomposed into memory modules (such as DRAM) and compute modules (such as CPUs), and connected with CXL Switch.
CXL Switch organizes all the memory modules into a memory pool and exposes it to the compute modules. 
The compute modules can dynamically request and release memory modules based on their demands, effectively improving the overall memory utilization efficiency in the data center. 
In the future, the data center will become a fabric composed of a large number of modular components, leads to \textbf{Composable Infrastructure}.

\subsection{Architecture of Pooling System}
In 2022, researchers from KAIST propose DirectCXL\cite{DirectCXL}, a software-hardware co-designed CXL memory pooling system. DirectCXL is primarily built upon CXL 2.0 and can be considered as one of the earliest publicly disclosed CXL memory pooling prototypes based on FPGA. In DirectCXL, a single CXL Switch connects multiple CXL memory modules and multiple processors to form a memory pool and computation pool. 
They reveal that CXL`s` latency mainly comes from PCIe bus, not related to the transmission size but CPU cache.

They expand DirectCXL by incorporating the features of CXL3.0 \cite{MemoryPoolingWithCXL}, upgrading it from a memory pool system to a more fine-grained and larger-scale composable system. 
They integrate the Multi-Header Logical Device (MH-LD) and Dynamic Capacity Device (DCD) features into DirectCXL, providing more fine-grained memory management and sharing capabilities. 
Additionally, they use a port-number-based routing mechanism to extend the CXL topology, allowing the integration of more CXL switches and devices, forming a large-scale, multi-level switch structure that can theoretically support up to 4096 devices and/or hosts.

Inspired by CXL, Mohammad Ewais et al. discuss the CXL based Disaggregated Datacenter (DDC)\cite{ddc}. 
They believe that CXL will help make the vision of a disaggregated datacenter a reality. They propose a system model with the following key elements: Compute Units, Caching, Memory Nodes, Virtual Memory, Interconnects and Networking, Storage, Management. They also discuss the limitations of existing research, such as the lack of a reasonable explanation for architectural choices, the convenience of design and adoption, latency issues, and the complexity of data sharing.

Unlike the physical memory pooling represented by DirectCXL, Emmanuel Amaro et al. from VMware propose a new memory disaggregation architecture based on Logical Memory Pools (LMPs) \cite{Amaro2023LogicalMP}. Compared to physical memory pooling, LMPs create memory pools by partitioning a portion of the local memory on each server, rather than using a dedicated memory device. The advantage of this approach is lower cost, support for near-memory computing without additional hardware, and more flexibility in design. The evaluations show that LMPs can execute workloads that are not feasible in physical memory pools, and their performance is also better due to faster access speeds.

Ming Liu proposes a new computing paradigm called \textbf{Fabric-Centric Computing} (FCC)\cite{FFC}. This approach addresses the challenges of emerging memory fabrics and composable infrastructures,  treating the memory fabric as a first-class citizen for instance, orchestration, and reclamation of computing resources. 
Liu discusses the differences between memory fabrics and communication fabrics, including the synchronous execution model, diverse memory node types, dependence on PCIe routing performance, fast execution engine context switching, and the introduction of passive fault domains. 
Based on these characteristics, Liu proposes the design principles of FCC, including data movement as a managed service, host-assisted memory node type-aware data structures, idempotent tasks and hardware-assisted scalable offloading, as well as a fabric-centric arbitrator implemented over dedicated channels.

\subsection{Interconnection in Pooling System}
In disaggregated systems, the distance between storage and computing resources (known as the Access Latency) often becomes a bottleneck to overall performance.

Andreas Geyer et al. discuss the various memory distances that may occur in a hardware-disaggregated environment, propose a classification of memory distances based on physical distance and transport layer\cite{Geyer2023WorkingWD}. They also discuss key use cases, including dynamic memory expansion, multi-socket server memory integration, shared memory pool across multiple servers, and the use of near-memory computing (PIM) and RDMA to reduce data transmission.

Andreas Geyer et al. also evaluate the impact of interconnect on disaggregated systems\cite{NeartoFar}. The paper focuses on discussing RDMA and CXL, which allow dynamic sharing of hardware resources between different servers. they systematically evaluated the impact of local Ultrapath Interconnect (UPI), RDMA over InfiniBand, and memory attached via CXL over PCIe on disaggregated systems. The results show that the performance of CXL is comparable to UPI and better than asynchronous RDMA.

Shu-Ting Wang et al. from UCSD, propose \textbf{Aurelia}\cite{Aurelia}, a new network architecture based on CXL. The current multi-level switching in CXL faces challenges in terms of scalability and latency. 
Aurelia addresses these challenges by architecting the CXL network with addressing, routing, and transport as network-level functions. 
Aurelia introduces a new addressing scheme that uses a 12-bit Fabric Node ID (FAN-ID) to enhance the port ID scheme in CXL, similar to the 32-bit IP addresses in IP networks. Additionally, Aurelia designs a routing protocol based on FAN-ID and an end-to-end congestion control mechanism to improve the scalability and reduce the latency of the CXL network. They also discuss the comparison of Aurelia with other technologies like NVLink and PCIe, as well as the cost-effectiveness and security implications of CXL.

The researchers from UCAS, propose \textbf{CXL over Ethernet} \cite{CXLoverEthernet}. By using RDMA to transmit CXL access requests, it overcomes the shortcomings of native CXL-based disaggregated systems, which are limited to the rack scale. The CXL memory requests can be directly sent to the FPGA, which then interacts with the remote memory through its network interface. This work demonstrates that it is possible to combine the advantages of RDMA and native CXL technologies.

Similar to the CXL over Ethernet, Zhonghua Wang et al. introduce a new low-latency and highly scalable memory pool system called \textbf{Rcmp}\cite{rcmp}. Rcmp achieves memory disaggregation by combining RDMA and CXL. Rcmp builds a CXL-based memory pool within a rack and connects different racks through RDMA to form a larger memory pool. To address the challenges of mismatched granularity, communication, and performance between RDMA and CXL, Rcmp proposes a global page-based memory space management, an efficient communication mechanism, a hot page identification and migration strategy, and an optimized RDMA RPC framework. 

The collaborative work between Ming Liu and Zeke Wang from Zhejiang University propose \textbf{rPCIeBench} \cite{rPCIeBench}, a hardware-software co-designed benchmark framework for systematically evaluating the performance of Routable PCIe (CXL), which is the key interconnect for building new composable infrastructures. 
They first analyzes the communication characteristics of the Routable PCIe and compares it with local PCIe. Then, they use rPCIeBench to delve into the traffic coordination behavior within the fabric, and propose three interesting findings: \textbf{Approximate Max-Min Bandwidth Partition}, \textbf{Fast End-to-End Bandwidth Synchronization}, and \textbf{Interference-free Orthogonal Paths}. Finally, they translate these insights into traffic coordination rules and develop an edge-constraint relaxation algorithm to estimate the transfer performance of PCIe flows on the shared fabric.

\subsection{Cost of Pooling System}
Researchers from Google\cite{ACaseAgainstCXL} have raised concerns about the potential application of CXL memory pooling in data centers and cloud systems. The paper points out three main issues that hinder the practicality of CXL memory pooling: cost, complexity, and utility. First, the cost of CXL memory pooling may outweigh the savings from reduced RAM. Second, the actual latency of CXL is higher than main memory, which requires extensive rewriting of networked applications, increasing software complexity. Finally, by analyzing production traces from Google and Azure Cloud, it is found that modern servers are large enough for most virtual machines, and simple virtual machine packing algorithms hardly leave any unused memory, which undermines the main motivation for memory pooling.

Contrary to the opinion of Google, Researchers from Microsoft argue that CXL-based memory pooling has significant cost advantages, and discuss some key design considerations for implementing a CXL-based disaggregated memory system in cloud computing scenarios\cite{pond2}. They analyze the critical design constraints for cloud service providers deploying CXL memory pools (memory pool access performance, compatibility with virtualization, memory pool system topology, etc.) as well as observations from actual deployments, and point out configuration examples with significant return on investment. They also compare the costs of memory pools with different pool sizes and CXL device types.





\section{Distributed Shared Memory}
\label{section:Distributed Shared Memory}

\begin{figure}[!t]
\centering
\includegraphics[width=0.49\textwidth]{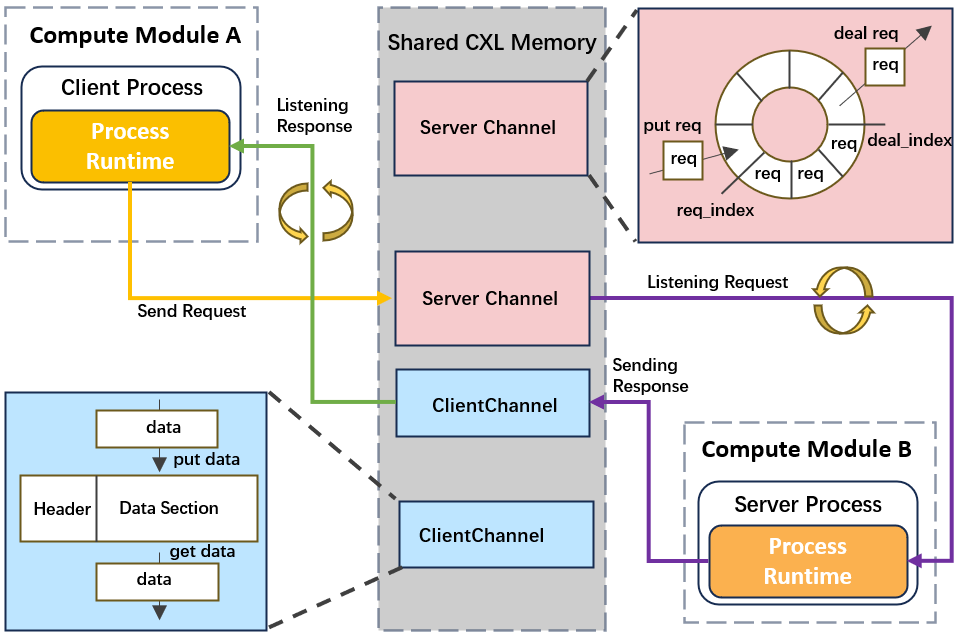}
\caption{Communication via Shared CXL Memory}
\label{sharing memory}
\end{figure}

Unlike memory pooling proposed in CXL2.0, CXL3.0 introduces \textbf{Memory Sharing}. A memory module is occupied by a single compute module in CXL2.0, whereas memory sharing allows a memory module to be shared by multiple compute modules, as shown in figure \ref{pooling memory}. 
CXL3.0 provides a hardware cache coherence mechanism, effectively avoids the major performance bottleneck of traditional distributed shared memory systems: the network-based cache coherence mechanism, and opens up new opportunities.

\subsection{Distributed Shared Memory with CXL 3.0}
Researchers from AMD discuss the design space of memory sharing using CXL technology \cite{jain2024memorysharingcxlhardware}. This work explores various methods to implement memory sharing under different CXL protocol versions (CXL 1.1, CXL 2.0, and CXL 3.0), including software-based shared memory system and hardware-level shared memory management mechanisms. This work also discusses the granularity and security issues of memory sharing and proposes corresponding solutions. 

Mingxing Zhang et al. propose CXL-SHM\cite{cxlshm}, an efficient distributed memory management system based on CXL. CXL-SHM leverages the performance advantages of CXL3.0 to improve the synchronization in DSM. Zhang evaluated the efficiency of CXL-SHM on real CXL hardware, CXL-SHM`s memory management performance is superior to the existing persistent memory allocators. CXL-SHM also demonstrates how to build efficient distributed applications by utilizing the shared CXL memory, such as a share-everything KVStore, proving its potential and flexibility in real-world applications.

Mingxing Zhang et al. also use shared CXL memory to provide memory elasticity for serverless computing.
They introduce \textbf{TrEnv} \cite{TrEnv}, a co-designed integration of the serverless platform and CXL memory pools. 
TrEnv introduces repurposable sandboxes, which can be shared across different functions, decreases the overhead of creating isolation sandboxes. 
Then, it augments the OS with "memory templates" that enable rapid restoration of function states stored on shared CXL memory. 

\subsection{Communication via Shared CXL Memory}
As shown in figure \ref{sharing memory}, the shared CXL memory enables efficient synchronisation and communication between distributed processes running on different compute modules.
The client process writes a request in the shared memory, and the server process polls the shared memory, detects the request, processes it, and sends response.

In \textbf{CXL-SHM}\cite{cxlshm}, Zhang et al. also propose a pass-by-reference RPC framework that improves communication efficiency by passing object references (rather than object values) in shared CXL memory.

On the basis of CXL-SHM\cite{cxlshm}, researchers from Alibaba propose \textbf{HydraRPC}\cite{HydraRPC}, an RPC system based on shared CXL memory. HydraRPC introduces a novel non-cached sharing mechanism. In data reading, it sets specific memory areas as uncachable to ensure that the data is transferred directly between CXL HDM and the processor, bypassing the CPU cache. In data writing, HydraRPC uses specific CPU instructions to ensure that data is written directly to the CXL HDM, rather than just to the local cache. This synchronizes non-temporal accesses using load/store fences, ensuring consistency. 

Similar to CXL-SHM\cite{cxlshm} and HydraRPC\cite{HydraRPC}, researchers from Intel and UCSC propose \textbf{NotNets}, 
a RPC system based on shared CXL memory\cite{NotNets}. NotNets stores message payloads and metadata on the distributed shared memory supported by CXL 3.0, and only passing references through shared memory to bypass the network. NotNets can be integrated into existing RPC frameworks, allowing host clusters with up to 512-1024 cores to transparently implement message passing semantics using the CXL attached memory pool, avoiding the main bottlenecks in the current RPC stack.

The researchers from UCSD, led by Steven Swanson propose RPCool\cite{rpcool}, an RPC system based on shared CXL memory. RPCool introduces an efficient access permission transfer mechanism to ensure the concurrent safety of payloads during the communication process. In addition, RPCool seamlessly integrates RDMA (responsible for cross-rack communication) and CXL-RPC (responsible for intra-rack communication), allowing its scale to be extended to the entire data center.

\subsection{Failure Tolerance}
\label{(Failure Tolerance}
While CXL supports multi-node cache-coherent shared memory, it introduces new failure types, such as data potentially being lost before a process, or a process potentially crashing before the data. However, the existing CXL shared memory lacks a failure model, making it challenging to understand and mitigate these failures.

In \textbf{CXL-SHM}\cite{cxlshm}, Zhang et al. use a special era-based non-blocking algorithm to maintain reference counting, effectively solving the synchronization, memory leaks, double release, and dangling pointer issues in distributed memory management. Even if some clients fail unexpectedly due to process crashes or machine failures, the system can automatically recover, ensuring stability and reliability.

Steven Swanson et al. from UCSD, firstly propose a failure model that classifies and handles \textbf{Data Failures} and \textbf{Process Failures} in CXL shared memory \cite{xu2024cxlsharedmemoryprogramming}. First, the researchers discuss data failures, which occur when the CXL memory device becomes unavailable. Next, they explore process failures and compare them to the failure model and programming solutions in persistent memory (PMEM). The researchers note that while PMEM shares some features with CXL, it does not fully address the volatility and low access latency challenges of CXL. This work also proposes customized solutions to the specific challenges of CXL systems.



\section{Future Research}
\label{Future Research}

Today, CXL has become a hot topic in both academic and industrial communities. However, in each research area, there are still significant gaps to be filled, or some challenges that have not yet been resolved. 
In this section, we will briefly discuss the gaps in existing research and the potential directions for future research.

\subsection{CXL Based Memory Extension}
\subsubsection{Memory Interleaving}
The Tiered Memory System(TMS) around CXL memory (as slow tier) and DRAM (as fast tier) is the most widely discussed topic, represented by TPP\cite{TPP,AccessLatencyisKey}. 
However, TMS only copes with latency issue, ignores the bandwidth issue. 
In TMS, all reads and writes are executed on the fast tier, which means, TMS`s overall bandwidth cannot exceed the upper bound of the fast tier. Once the bandwidth of the memory is saturated \cite{AccessLatencyisKey}, the performance of the workloads degrades.

One possible solution is \textbf{Memory Interleaving}\cite{liu2024exploring,tang2024exploring}, where CXL memory is not used as a slow tier, but is configured in parallel with local DRAM. Workloads can access both CXL memory and local memory directly. 
Because CXL is built on PCIe, requests to access CXL memory do not consume DRAM bandwidth.
In memory interleaving system, the overall bandwidth can be the sum of CXL memory and DRAM in theory, which shows great potential in bandwidth.


\subsubsection{Function Offloading}
Since CXL Memory requires dedicated hardware, offloading workload`s specific function to hardware is a promising approach to avoid high-latency data movement. 
However, existing works focus too much on optimizing for specific workloads. For other workloads, these optimizations may not help or even degrade performance, which lacks flexibility and increases the cost of infrastructure. 
There is an urgent need for a generic, workload-decoupled function offloading solution that can ensure performance improvement while also providing flexibility. 
One possible solution is to integrate some general-purpose computing units (similar to NDP cores) into the CXL devices, enabling them to execute custom near-data computing tasks.


\subsection{CXL Based Unified Memory}
CXL-based unified memory may be the most promising research area, as a unified, cache-coherent memory domain is crucial for heterogeneous computing systems, which effectively improve the collaborative efficiency.
\subsubsection{GPU Memory Extension}

Apart from the classic solution of offloading GPU data to CPU-attached DRAM, CXL memory provides a new design space. However, since existing commercial GPUs do not provide hardware support, for GPUs to offload data to CXL memory, they must go through CPU as a relay, which brings new challenges: first, the longer access path of GPU-CPU-CXL will lead to higher latency; second, going through the CPU not only occupies the limited bandwidth between CPU and GPU, but also incurs additional CPU overhead.


\subsubsection{Heterogeneous Device Collaboration}

The collaboration between heterogeneous accelerators (e.g. GPU and NIC) is a promising research area. PCIe devices are often divided into separate memory domains and P2P data access is not available. 
CXL.cache allows devices to directly access data in each other's memory, greatly improving the efficiency of device collaboration.
However, 
there are too few accelerators that support CXL.cache, lacking real use case.

\subsection{CXL Based Independent Memory}
CXL memory is located on a separate device, which has independent power supply. 
This is both a challenge and an opportunity. 
As for the challenge, we have introduced the new fault tolerance model in section \ref{(Failure Tolerance}. 
As for the opportunity, CXL memory is \textbf{independent} of host, the status of the host or process does not affect the CXL memory, which gives us great research space.



\subsubsection{Stateful Workloads Upgrade or Migration}
During migration, instances go through shutdowns, upgrades, and reboots. They can not respond to any requests until the migration finished. This period of time is called downtime.
For stateless workloads, downtime can be reduced by starting new instances early and switching over request traffic while new instance getting ready.
However, for stateful workloads, the data in old instance is continuously modified, new instance need to continuous sync data from old one until the new one catches up, leads to a long-term process.

With independent CXL memory, the old instance can place the stateful data on shared CXL memory rather than its own memory. After the new instance starts, the old instance can be stopped directly, switching the request traffic. The new instance can directly operate the data in the CXL memory without any synchronization. This approach decouples host status and data, making the host stateless.

\subsubsection{Workloads Cold Start}

Cold startup is an important challenge in serverless computing, as loading container images from disk or remote storage into memory introduces great overhead to the startup process. 
With CXL memory, researchers can maintain a \textbf{hot image pool}. All hosts connected to the CXL memory do not need to access remote storage, but can directly pull the pre-warmed images from the CXL memory with extremely low latency.

\subsubsection{CXL File System}

Similar to the existing research on in-memory file systems, it is also possible to migrate the file system to the CXL memory, taking advantage of the lower latency. However, the fault tolerance of CXL memory (data failure) needs to be considered.
\subsection{CXL Based Disaggregated System}
There is a large gap between existing research and practical CXL-based disaggregated system. Existing research is still theoretical, trying to establish a basic model for composable infrastructure. The interconnection, management, scalability, cost of composable infrastructure are still blank areas.
\subsubsection{CXL Switch}
The major part of CXL-based disaggregated system is \textbf{CXL Switch}, which is still in the prototype stage.
CXL Switch must address following challenges.
Firstly, the \textbf{latency}. In a larger-scale cascaded switching system, the end-to-end latency will be several times higher than the direct latency when the host CPU is directly connected to the CXL device. 
Secondly, the \textbf{scalability}. Connecting as many devices and CPUs as possible while maintaining their cache coherence.

\subsubsection{System Model}
The disaggregated system originates from distributed system, consists of memory devices and processors, but has a larger design space, and faces more challenges.
It's necessary to establish a system model to describe the system components and functions, such as control plane, computing plane, data plane and interconnection.


\subsubsection{System Cost}
For cloud providers and data center managers, cost is the most important factor. Existing research focus too much on optimizing CXL system`s performance, ignoring the economic benefits that CXL can bring. 
The industry urgently needs a quantitative model to describe the memory utilization improvement by CXL, and whether the improvement can cover the hardware costs introduced by CXL.

\subsubsection{Virtualization}
Cloud providers also urgently need a virtualization solution to commercialize CXL \cite{pond2,CXLINVIRTUAL}. Compared to the traditional VM sales pattern (where the ratio of CPU and memory is fixed), CXL VM instances deployed on the memory pool can configure memory flexibly. However, the efficiency of VM`s autoscaler, the data reliability and the consistency during the scaling process, are still open questions. 

\subsubsection{Cross-Rack Interconnect}
In addition to CXL switch cascade, another option for scaling up the system is to interconnect the racks via fast networks (e.g. RDMA). 
However, there is a mismatch in semantics, granularity, and latency between RDMA and CXL, making it difficult to directly forward memory requests, further research is needed.




\subsection{CXL Based Distributed Shared Memory}
Due to the unavailability of CXL3.0 hardware, research on CXL-based distributed shared memory is still at the demo stage, lacking real-world use cases. 
\subsubsection{Security}
The most important thing in shared CXL memory is security (or access control). All processors connected to shared CXL device are able to access the shared memory area freely, posing a large attack surface. 
Researchers need to design a high-performance memory permission mechanism like Intel Memory Protection Keys\cite{mpk} to ensure the security.

\subsubsection{Scalability}
The scalability of shared memory is also a concern. There are two factors that matters: \textbf{Bandwidth} and \textbf{Cache Coherence}. 
Firstly, in distributed systems, the scale of memory requests will be much larger, the current bandwidth of CXL memory will saturate quickly.
Secondly, as AMD reports\cite{jain2024memorysharingcxlhardware}, maintaining large-scale CPU`s cache coherence is difficult. As the size of shared memory grows, a precise snoop filter that tracks every 64B cache line may turn out to be an impractical solution.

\subsubsection{Application}
\textbf{Cross-Node Communication} and \textbf{Share-Everything Database} are most attractive application of CXL distributed shared memory. 
There's still work to do. 
Shared memory communication still needs a complete protocol and congestion control mechanism.
Databases also face the challenge of large-scale concurrent accesses.





\subsubsection{Failure Tolerance}
CXL supports cross-host memory sharing, but it also introduces new types of failures. These failures include: data loss before a process ends, or processes crashing while data still available. Researchers believe that existing CXL shared memory lacks a failure model \cite{xu2024cxlsharedmemoryprogramming}, making it challenging to understand and mitigate these failures.

\section{Memory-Centric Computing in CXL Fabric}
\label{Memory-Centric Computing}

Memory-centric computing aims to enable computation capability in and near all places where data is generated and stored\cite{mutlu2023memorycentriccomputing,Mcc1}. 

CXL brings new opportunities, it 
makes computing and memory resources in the data center no longer tightly coupled together. Interconnected via CXL fabric, the data centers look like a hyperscale node, moving towards the \textbf{Composable Infrastructure}. In such hyperscale nodes, the distance between data and computating, and the data movement, need to be reconsidered. 
In the section, we propose three design points.

\textbf{\textit{The Distance of Movement:}} In single-machine computing systems, distance refers to the overhead of data movement between memory and CPU's 3 level caches or registers. 
However, in CXL fabric (a multilevel spine-leaf CXL switching system), the smallest unit is no longer a single machine, it is transformed into a \textbf{CELL} consisting of the leaf switch and the devices directly connected to it. 
While CXL's cross-cell data movement overhead is much lower than network, it is still greater than the data movement overhead in the same cell, as shown in pond\cite{pond} (155ns vs. 270ns). 

We define 2 distance, \textbf{Intra-Cell} and \textbf{Inter-Cell} distance. For intra-cell, we can offload function on memory devices to avoid data movement, as same as the existing works in section \ref{NMP}. For inter-cell, we schedule computing devices according to the location of the data, prioritizing the computing devices within the same cell to reduce the distance.


\textbf{\textit{The Data Movement:}}
Besides the data movement between CPU and memory, data movement between cross-node CPUs receives greater attention(e.g. network).
In CXL fabric, shared memory will be more efficient for coordination
between multiple computing devices. Its throughput will be directly dependent on the memory bandwidth and the maximum number of Load/Store instructions that can be accommodated.


Data movement in shared memory is associated with two entities: senders and receivers. To reduce the overhead, they should be decoupled, which means, the movement request is triggered by the sender, but executed by the \textbf{Movement Executor}, which can be the sender, the receiver, or a third-party service. 
The executor should effectively use synchronous and asynchronous execution. Synchronous execution serves latency-sensitive requests or those closely coupled with the current execution context (e.g., traversing data structures). The executor can enhance the efficiency use caching and prefetching to take advantage of data locality. 
For asynchronous execution, after a movement request is initiated, senders will start other task. Once the movement is finished, the movement executor will give senders or receivers a signal to notify.


\textbf{\textit{The Memory Orchestration:}}
In CXL fabric, different computing tasks have vastly different memory usage patterns. 
Heterogeneous memory devices(e.g,DRAM, NVMe) also have vastly different properties.
Orchestration over heterogeneous tasks and memory is quite difficult. 

Data structures could be a good abstraction to describe memory usage patterns. They can be designed for different computing tasks with different characteristics. 

The runtime system for cxl fabric should have the following capabilities.
First, it is able to recognize the memory usage patterns via data structures or object temperature.
Then, it is able to allocate appropriate memory resources to best fit the usage patterns.
Lastly, it is able migrate data between heterogeneous memory to achieve dynamic adjustment.
\section{Conclusion}
\label{conclusion}

This paper studies the CXL based next-generation computing systems from single-machine to distributed. 

In single-machine computing systems, this paper classify the existing research into two categories according to the interconnection types: \textbf{Memory Expansion} and \textbf{Unified Memory}. 
Memory Expansion focuses on processors and memory devices, attempting to address memory wall with tiered memory system and near-memory processing. 
Unified Memory focuses on processors and accelerators, enabling more efficient cooperation in heterogeneous computing systems. 

In distributed computing systems, this paper discusses how to build high-performance disaggregation systems based on CXL infrastructure, enabling flexible and efficient resource utilization. 
In addition, this paper also talks about distributed shared CXL memory, which improves the efficiency of communication and synchronization in distributed systems.

Finally, this paper discusses future research trends in CXL-based computing systems and envision a data-centric computing paradigm. 
We hope this paper provides newcomers with a general understanding and offers fresh inspiration for researchers in the CXL community.


\ifCLASSOPTIONcaptionsoff
  \newpage
\fi



\bibliographystyle{bibtex/IEEEtran}
\bibliography{bibtex/IEEEabrv,reference}

\begin{thebibliography}{10}
\providecommand{\url}[1]{#1}
\csname url@samestyle\endcsname
\providecommand{\newblock}{\relax}
\providecommand{\bibinfo}[2]{#2}
\providecommand{\BIBentrySTDinterwordspacing}{\spaceskip=0pt\relax}
\providecommand{\BIBentryALTinterwordstretchfactor}{4}
\providecommand{\BIBentryALTinterwordspacing}{\spaceskip=\fontdimen2\font plus
\BIBentryALTinterwordstretchfactor\fontdimen3\font minus \fontdimen4\font\relax}
\providecommand{\BIBforeignlanguage}[2]{{%
\expandafter\ifx\csname l@#1\endcsname\relax
\typeout{** WARNING: IEEEtran.bst: No hyphenation pattern has been}%
\typeout{** loaded for the language `#1'. Using the pattern for}%
\typeout{** the default language instead.}%
\else
\language=\csname l@#1\endcsname
\fi
#2}}
\providecommand{\BIBdecl}{\relax}
\BIBdecl

\bibitem{memorywall}
A.~Gholami, Z.~Yao, S.~Kim, C.~Hooper, M.~W. Mahoney, and K.~Keutzer, ``Ai and memory wall,'' \emph{IEEE Micro}, vol.~44, no.~3, pp. 33--39, 2024.

\bibitem{intro_to_cxl}
D.~Das~Sharma, R.~Blankenship, and D.~Berger, ``An introduction to the compute express link (cxl) interconnect,'' \emph{ACM Comput. Surv.}, vol.~56, no.~11, Jul. 2024.

\bibitem{Disaggregation_survey}
\BIBentryALTinterwordspacing
H.~Al~Maruf and M.~Chowdhury, ``Memory disaggregation: Advances and open challenges,'' \emph{SIGOPS Oper. Syst. Rev.}, vol.~57, no.~1, p. 29–37, Jun. 2023. [Online]. Available: \url{https://doi.org/10.1145/3606557.3606562}
\BIBentrySTDinterwordspacing

\bibitem{DSM_survey}
\BIBentryALTinterwordspacing
B.~Nitzberg and V.~Lo, ``Distributed shared memory: A survey of issues and algorithms,'' \emph{Computer}, vol.~24, no.~8, p. 52–60, Aug. 1991. [Online]. Available: \url{https://doi.org/10.1109/2.84877}
\BIBentrySTDinterwordspacing

\bibitem{cxl1}
\BIBentryALTinterwordspacing
``Compute express link 3.0,'' 2022. [Online]. Available: \url{https://www.computeexpresslink.org/_files/ugd/0c1418_a8713008916044ae9604405d10a7773b.pdf}
\BIBentrySTDinterwordspacing

\bibitem{cxl2}
\BIBentryALTinterwordspacing
``Compute express link cxl 3.0 is the exciting building block for disaggregation.'' 2022. [Online]. Available: \url{https://www.servethehome.com/compute-expresslink-cxl-3-0-is-the-exciting-building-block-for-disaggregation/}
\BIBentrySTDinterwordspacing

\bibitem{cxl3}
\BIBentryALTinterwordspacing
``Compute express link™: The breakthrough cpu-to-device interconnect.'' 2022. [Online]. Available: \url{https://www.computeexpresslink.org/home}
\BIBentrySTDinterwordspacing

\bibitem{cxlintro2}
D.~D. Sharma, ``Compute express link (cxl): Enabling heterogeneous data-centric computing with heterogeneous memory hierarchy,'' \emph{IEEE Micro}, vol.~43, no.~2, pp. 99--109, 2023.

\bibitem{TPP}
H.~Maruf, H.~Wang, A.~Dhanotia, J.~Weiner, N.~Agarwal, P.~Bhattacharya, C.~Petersen, M.~Chowdhury, S.~Kanaujia, and P.~Chauhan, ``\BIBforeignlanguage{en-US}{Tpp: Transparent page placement for cxl-enabled tiered-memory},'' \emph{\BIBforeignlanguage{en-US}{ASPLOS}}, Jan 2023.

\bibitem{AccessLatencyisKey}
M.~Vuppalapati and R.~Agarwal, ``Tiered memory management: Access latency is the key!'' in \emph{Proceedings of the ACM SIGOPS 30th Symposium on Operating Systems Principles}, ser. SOSP '24.\hskip 1em plus 0.5em minus 0.4em\relax New York, NY, USA: Association for Computing Machinery, 2024, p. 79–94.

\bibitem{CXL-ANNS}
J.~Jang, H.~Choi, H.~Bae, S.~Lee, M.~Kwon, and M.~Jung, ``{CXL-ANNS}: {Software-Hardware} collaborative memory disaggregation and computation for {Billion-Scale} approximate nearest neighbor search,'' in \emph{2023 USENIX Annual Technical Conference (USENIX ATC 23)}.\hskip 1em plus 0.5em minus 0.4em\relax Boston, MA: USENIX Association, Jul. 2023, pp. 585--600.

\bibitem{CXL-ANNS2}
{J. Jang, H. Choi, H. Bae, S. Lee, M. Kwon, and M. Jung}, ``Bridging software-hardware for cxl memory disaggregation in billion-scale nearest neighbor search,'' \emph{ACM Trans. Storage}, vol.~20, no.~2, Feb. 2024.

\bibitem{NeoMem}
Z.~Zhou, Y.~Chen, T.~Zhang, Y.~Wang, R.~Shu, S.~Xu, P.~Cheng, L.~Qu, Y.~Xiong, J.~Zhang, and G.~Sun, ``Neomem: Hardware/software co-design for cxl-native memory tiering,'' in \emph{2024 57th IEEE/ACM International Symposium on Microarchitecture (MICRO)}, 2024.

\bibitem{recxl}
H.~Liu, L.~Zheng, Y.~Huang, J.~Zhou, C.~Liu, R.~Wang, X.~Liao, H.~Jin, and J.~Xue, ``Enabling efficient large recommendation model training with near cxl memory processing,'' in \emph{2024 ACM/IEEE 51st Annual International Symposium on Computer Architecture (ISCA)}, 2024.

\bibitem{BreakingBarriers}
D.~Gouk, S.~Kang, H.~Bae, E.~Ryu, S.~Lee, D.~Kim, J.~Jang, and M.~Jung, ``Breaking barriers: Expanding gpu memory with sub-two digit nanosecond latency cxl controller,'' in \emph{Proceedings of the 16th ACM Workshop on Hot Topics in Storage and File Systems}, ser. HotStorage '24.\hskip 1em plus 0.5em minus 0.4em\relax New York, NY, USA: Association for Computing Machinery, 2024, p. 108–115.

\bibitem{Kwon2023FailureTT}
M.~Kwon, J.~Jang, H.~Choi, S.~Lee, and M.~Jung, ``Failure tolerant training with persistent memory disaggregation over cxl,'' \emph{IEEE Micro}, vol.~43, pp. 66--75, 2023.

\bibitem{GPUGraphProcessing}
S.~Sano, Y.~Bando, K.~Hiwada, H.~Kajihara, T.~Suzuki, Y.~Nakanishi, D.~Taki, A.~Kaneko, and T.~Shiozawa, ``Gpu graph processing on cxl-based microsecond-latency external memory,'' in \emph{Proceedings of the SC '23 Workshops of The International Conference on High Performance Computing, Network, Storage, and Analysis}, ser. SC-W '23.\hskip 1em plus 0.5em minus 0.4em\relax New York, NY, USA: Association for Computing Machinery, 2023, p. 962–972.

\bibitem{pond}
H.~Li, D.~Berger, S.~Novakovic, L.~Hsu, D.~Ernst, P.~Zardoshti, M.~Shah, S.~Rajadnya, S.~Lee, I.~Agarwal, M.~Hill, M.~Fontoura, and R.~Bianchini, ``\BIBforeignlanguage{en-US}{Pond: Cxl-based memory pooling systems for cloud platforms},'' \emph{\BIBforeignlanguage{en-US}{ASPLOS}}, Jan 2023.

\bibitem{DirectCXL}
D.~Gouk, S.~Lee, M.~Kwon, and M.~Jung, ``\BIBforeignlanguage{en-US}{Direct access, high-performance memory disaggregation with directcxl},'' \emph{\BIBforeignlanguage{en-US}{ATC}}, Jan 2022.

\bibitem{ddc}
M.~Ewais and P.~Chow, ``Ddc: A vision for a disaggregated datacenter,'' 2024.

\bibitem{NotNets}
P.~Alvaro, M.~Adiletta, A.~Cockroft, F.~Hady, R.~Illikkal, E.~Ramos, J.~Tsai, and R.~Soulé, ``Notnets: Accelerating microservices by bypassing the network,'' 2024.

\bibitem{cxlshm}
M.~Zhang, T.~Ma, J.~Hua, Z.~Liu, K.~Chen, N.~Ding, F.~Du, J.~Jiang, T.~Ma, and Y.~Wu, ``Partial failure resilient memory management system for (cxl-based) distributed shared memory,'' in \emph{Proceedings of the 29th Symposium on Operating Systems Principles}, ser. SOSP '23.\hskip 1em plus 0.5em minus 0.4em\relax New York, NY, USA: Association for Computing Machinery, 2023, p. 658–674.

\bibitem{redis}
\BIBentryALTinterwordspacing
``Redis ltd.'' 2024. [Online]. Available: \url{https://redis.io/.}
\BIBentrySTDinterwordspacing

\bibitem{sun2023demystifying}
Y.~Sun, Y.~Yuan, Z.~Yu, R.~Kuper, C.~Song, J.~Huang, H.~Ji, S.~Agarwal, J.~Lou, I.~Jeong \emph{et~al.}, ``Demystifying cxl memory with genuine cxl-ready systems and devices,'' in \emph{Proceedings of the 56th Annual IEEE/ACM International Symposium on Microarchitecture}, 2023, pp. 105--121.

\bibitem{tang2024exploring}
Y.~Tang, P.~Zhou, W.~Zhang, H.~Hu, Q.~Yang, H.~Xiang, T.~Liu, J.~Shan, R.~Huang, C.~Zhao \emph{et~al.}, ``Exploring performance and cost optimization with asic-based cxl memory,'' in \emph{Proceedings of the Nineteenth European Conference on Computer Systems}, 2024, pp. 818--833.

\bibitem{liu2024exploring}
J.~Liu, X.~Wang, J.~Wu, S.~Yang, J.~Ren, B.~Shankar, and D.~Li, ``Exploring and evaluating real-world cxl: Use cases and system adoption,'' \emph{arXiv preprint arXiv:2405.14209}, 2024.

\bibitem{wahlgren2022evaluating}
J.~Wahlgren, M.~Gokhale, and I.~B. Peng, ``Evaluating emerging cxl-enabled memory pooling for hpc systems,'' in \emph{2022 IEEE/ACM Workshop on Memory Centric High Performance Computing (MCHPC)}.\hskip 1em plus 0.5em minus 0.4em\relax IEEE, 2022, pp. 11--20.

\bibitem{ZeRO-Offload}
J.~Ren, S.~Rajbhandari, R.~Aminabadi, O.~Ruwase, S.~Yang, M.~Zhang, L.~Dong, and Y.~He, ``\BIBforeignlanguage{en-US}{Zero-offload: Democratizing billion-scale model training},'' \emph{\BIBforeignlanguage{en-US}{Cornell University - arXiv,Cornell University - arXiv}}, Jan 2021.

\bibitem{FlexGen}
Y.~Sheng, L.~Zheng, B.~Yuan, Z.~Li, M.~Ryabinin, B.~Chen, P.~Liang, C.~Zhang, I.~Stoica, and C.~Ré, ``\BIBforeignlanguage{en-US}{High-throughput generative inference of large language model with a single gpu}.''

\bibitem{cxlHybridPool}
Q.~Yang, R.~Jin, B.~Davis, D.~Inupakutika, and M.~Zhao, ``Performance evaluation on cxl-enabled hybrid memory pool,'' in \emph{2022 IEEE International Conference on Networking, Architecture and Storage (NAS)}, 2022, pp. 1--5.

\bibitem{yang2023cxlmemsimpuresoftwaresimulated}
Y.~Yang, P.~Safayenikoo, J.~Ma, T.~A. Khan, and A.~Quinn, ``Cxlmemsim: A pure software simulated cxl.mem for performance characterization,'' 2023.

\bibitem{gond2024emucxlemulationframeworkcxlbased}
R.~Gond and P.~Kulkarni, ``emucxl: an emulation framework for cxl-based disaggregated memory applications,'' 2024.

\bibitem{mess}
P.~Esmaili-Dokht, F.~Sgherzi, V.~S. Girelli, I.~Boixaderas, M.~Carmin, A.~Monemi, A.~Armejach, E.~Mercadal, G.~Llort, P.~Radojković, M.~Moreto, J.~Giménez, X.~Martorell, E.~Ayguadé, J.~Labarta, E.~Confalonieri, R.~Dubey, and J.~Adlard, ``A mess of memory system benchmarking, simulation and application profiling,'' in \emph{2024 57th IEEE/ACM International Symposium on Microarchitecture (MICRO)}, 2024.

\bibitem{DRackSim}
A.~Puri, K.~Bellamkonda, K.~Narreddy, J.~Jose, V.~Tamarapalli, and V.~Narayanan, ``Dracksim: Simulating cxl-enabled large-scale disaggregated memory systems,'' in \emph{Proceedings of the 38th ACM SIGSIM Conference on Principles of Advanced Discrete Simulation}, ser. SIGSIM-PADS '24.\hskip 1em plus 0.5em minus 0.4em\relax New York, NY, USA: Association for Computing Machinery, 2024, p. 3–14.

\bibitem{Nomad}
\BIBentryALTinterwordspacing
L.~Xiang, Z.~Lin, W.~Deng, H.~Lu, J.~Rao, Y.~Yuan, and R.~Wang, ``Nomad: {Non-Exclusive} memory tiering via transactional page migration,'' in \emph{18th USENIX Symposium on Operating Systems Design and Implementation (OSDI 24)}.\hskip 1em plus 0.5em minus 0.4em\relax Santa Clara, CA: USENIX Association, Jul. 2024, pp. 19--35. [Online]. Available: \url{https://www.usenix.org/conference/osdi24/presentation/xiang}
\BIBentrySTDinterwordspacing

\bibitem{SMT}
K.~Kim, H.~Kim, J.~So, W.~Lee, J.~Im, S.~Park, J.~Cho, and H.~Song, ``Smt: Software-defined memory tiering for heterogeneous computing systems with cxl memory expander,'' \emph{IEEE Micro}, vol.~43, no.~2, pp. 20--29, 2023.

\bibitem{song2023lightweightfrequencybasedtieringcxl}
\BIBentryALTinterwordspacing
K.~Song, J.~Yang, S.~Liu, and G.~Pekhimenko, ``Lightweight frequency-based tiering for cxl memory systems,'' 2023. [Online]. Available: \url{https://arxiv.org/abs/2312.04789}
\BIBentrySTDinterwordspacing

\bibitem{tirumalasetty2024exploringdramcacheprefetching}
\BIBentryALTinterwordspacing
C.~Tirumalasetty and N.~Annapreddy, ``Exploring dram cache prefetching for pooled memory,'' 2024. [Online]. Available: \url{https://arxiv.org/abs/2406.14778}
\BIBentrySTDinterwordspacing

\bibitem{cxlssd}
S.-P. Yang, M.~Kim, S.~Nam, J.~Park, J.~yong Choi, E.~H. Nam, E.~Lee, S.~Lee, and B.~S. Kim, ``Overcoming the memory wall with {CXL-Enabled} {SSDs},'' in \emph{2023 USENIX Annual Technical Conference (USENIX ATC 23)}.\hskip 1em plus 0.5em minus 0.4em\relax Boston, MA: USENIX Association, Jul. 2023, pp. 601--617.

\bibitem{CacheinHand}
M.~Kwon, S.~Lee, and M.~Jung, ``Cache in hand: Expander-driven cxl prefetcher for next generation cxl-ssd,'' in \emph{Proceedings of the 15th ACM Workshop on Hot Topics in Storage and File Systems}, ser. HotStorage '23.\hskip 1em plus 0.5em minus 0.4em\relax New York, NY, USA: Association for Computing Machinery, 2023, p. 24–30.

\bibitem{Hellobytes}
M.~Jung, ``Hello bytes, bye blocks: Pcie storage meets compute express link for memory expansion (cxl-ssd),'' in \emph{Proceedings of the 14th ACM Workshop on Hot Topics in Storage and File Systems}, ser. HotStorage '22.\hskip 1em plus 0.5em minus 0.4em\relax New York, NY, USA: Association for Computing Machinery, 2022, p. 45–51.

\bibitem{DeepMemoryDL}
\BIBentryALTinterwordspacing
M.~Arif, K.~Assogba, M.~M. Rafique, and S.~Vazhkudai, ``Exploiting cxl-based memory for distributed deep learning,'' in \emph{Proceedings of the 51st International Conference on Parallel Processing}, ser. ICPP '22.\hskip 1em plus 0.5em minus 0.4em\relax New York, NY, USA: Association for Computing Machinery, 2023. [Online]. Available: \url{https://doi.org/10.1145/3545008.3545054}
\BIBentrySTDinterwordspacing

\bibitem{CXLKVLLM}
Y.~T. ea~tl., ``Exploring cxl-based kv cache storage for llmserving,'' in \emph{Machine Learning for Systems Workshop at NeurIPS 2024}, 2024.

\bibitem{clay}
S.~Yun, H.~Nam, K.~Kyung, J.~Park, B.~Kim, Y.~Kwon, E.~Lee, and J.~H. Ahn, ``Clay: Cxl-based scalable ndp architecture accelerating embedding layers,'' in \emph{Proceedings of the 38th ACM International Conference on Supercomputing}, ser. ICS '24.\hskip 1em plus 0.5em minus 0.4em\relax New York, NY, USA: Association for Computing Machinery, 2024, p. 338–351.

\bibitem{BEACON}
W.~Huangfu, K.~T. Malladi, A.~Chang, and Y.~Xie, ``Beacon: Scalable near-data-processing accelerators for genome analysis near memory pool with the cxl support,'' in \emph{Proceedings of the 55th Annual IEEE/ACM International Symposium on Microarchitecture}, ser. MICRO '22.\hskip 1em plus 0.5em minus 0.4em\relax IEEE Press, 2023, p. 727–743.

\bibitem{CXL-PNM}
S.-S. Park, K.~Kim, J.~So, J.~Jung, J.~Lee, K.~Woo, N.~Kim, Y.~Lee, H.~Kim, Y.~Kwon, J.~Kim, J.~Lee, Y.~Cho, Y.~Tai, J.~Cho, H.~Song, J.~H. Ahn, and N.~S. Kim, ``An lpddr-based cxl-pnm platform for tco-efficient inference of transformer-based large language models,'' in \emph{2024 IEEE International Symposium on High-Performance Computer Architecture (HPCA)}, 2024, pp. 970--982.

\bibitem{cms}
J.~Sim, S.~Ahn, T.~Ahn, S.~Lee, M.~Rhee, J.~Kim, K.~Shin, D.~Moon, E.~Kim, and K.~Park, ``Computational cxl-memory solution for accelerating memory-intensive applications,'' \emph{IEEE Computer Architecture Letters}, vol.~22, no.~1, pp. 5--8, 2023.

\bibitem{Polaris}
Z.~Zhou, S.~Xu, Y.~Chen, T.~Zhang, R.~Shu, L.~Qu, P.~Cheng, Y.~Xiong, and G.~Sun, ``Polaris: Enhancing cxl-based memory expanders with memory-side prefetching,'' in \emph{Advanced Parallel Processing Technologies: 15th International Symposium, APPT 2023, Nanchang, China, August 4–6, 2023, Proceedings}.\hskip 1em plus 0.5em minus 0.4em\relax Berlin, Heidelberg: Springer-Verlag, 2023, p. 19–39.

\bibitem{M2NDP}
H.~Ham, J.~Hong, G.~Park, Y.~Shin, O.~Woo, W.~Yang, J.~Bae, E.~Park, H.~Sung, E.~Lim, and G.~Kim, ``Low-overhead general-purpose near-data processing in cxl memory expanders,'' in \emph{2024 57th IEEE/ACM International Symposium on Microarchitecture (MICRO)}, 2024.

\bibitem{UDON}
J.~Hermes, J.~Minor, M.~Wu, A.~Patil, and E.~V. Hensbergen, ``Udon: A case for offloading to general purpose compute on cxl memory,'' \emph{ArXiv}, vol. abs/2404.02868, 2024.

\bibitem{LMB}
J.~Wang, X.~Zhang, C.~Tang, X.~Chen, and T.~Lu, ``Lmb: Augmenting pcie devices with cxl-linked memory buffer,'' 2024.

\bibitem{DesignandanalysisCXL}
A.~M. Cabrera, A.~R. Young, and J.~S. Vetter, ``Design and analysis of cxl performance models for tightly-coupled heterogeneous computing,'' in \emph{Proceedings of the 1st International Workshop on Extreme Heterogeneity Solutions}, ser. ExHET '22.\hskip 1em plus 0.5em minus 0.4em\relax New York, NY, USA: Association for Computing Machinery, 2022.

\bibitem{SynergizingCXLwithUnified}
J.~Lee and J.~Kim, ``Synergizing cxl with unified memory for scalable gpu memory expansion,'' in \emph{2024 International Conference on Electronics, Information, and Communication (ICEIC)}, 2024, pp. 1--4.

\bibitem{Arif2023AcceleratingPO}
M.~Arif, A.~Maurya, and M.~M. Rafique, ``Accelerating performance of gpu-based workloads using cxl,'' \emph{Proceedings of the 13th Workshop on AI and Scientific Computing at Scale using Flexible Computing}, 2023.

\bibitem{salus}
R.~Abdullah, H.~Lee, H.~Zhou, and A.~Awad, ``Salus: Efficient security support for cxl-expanded gpu memory,'' \emph{2024 IEEE International Symposium on High-Performance Computer Architecture (HPCA)}, 2024.

\bibitem{MemoryPoolingWithCXL}
D.~Gouk, M.~Kwon, H.~Bae, S.~Lee, and M.~Jung, ``Memory pooling with cxl,'' \emph{IEEE Micro}, vol.~43, no.~2, pp. 48--57, 2023.

\bibitem{Amaro2023LogicalMP}
E.~Amaro, S.~Wang, A.~Panda, and M.~K. Aguilera, ``Logical memory pools: Flexible and local disaggregated memory,'' \emph{Proceedings of the 22nd ACM Workshop on Hot Topics in Networks}, 2023.

\bibitem{FFC}
M.~Liu, ``Fabric-centric computing,'' in \emph{Proceedings of the 19th Workshop on Hot Topics in Operating Systems}, ser. HOTOS '23.\hskip 1em plus 0.5em minus 0.4em\relax New York, NY, USA: Association for Computing Machinery, 2023, p. 118–126.

\bibitem{Geyer2023WorkingWD}
A.~M. Geyer, D.~Ritter, D.~H. Lee, M.~Ahn, J.~Pietrzyk, A.~Krause, D.~Habich, and W.~Lehner, ``Working with disaggregated systems. what are the challenges and opportunities of rdma and cxl?'' in \emph{Datenbanksysteme f{\"u}r Business, Technologie und Web}, 2023.

\bibitem{NeartoFar}
A.~Geyer, J.~Pietrzyk, A.~Krause, D.~Habich, W.~Lehner, C.~F\"{a}rber, and T.~Willhalm, ``Near to far: An evaluation of disaggregated memory for in-memory data processing,'' in \emph{Proceedings of the 1st Workshop on Disruptive Memory Systems}, ser. DIMES '23.\hskip 1em plus 0.5em minus 0.4em\relax New York, NY, USA: Association for Computing Machinery, 2023, p. 16–22.

\bibitem{Aurelia}
S.-T. Wang and W.~Wang, ``Aurelia: Cxl fabric with tentacle,'' \emph{Proceedings of the 4th Workshop on Resource Disaggregation and Serverless}, 2023.

\bibitem{CXLoverEthernet}
C.~Wang, K.~He, R.~Fan, X.~Wang, W.~Wang, and Q.~Hao, ``Cxl over ethernet: A novel fpga-based memory disaggregation design in data centers,'' in \emph{2023 IEEE 31st Annual International Symposium on Field-Programmable Custom Computing Machines (FCCM)}, 2023, pp. 75--82.

\bibitem{rcmp}
Z.~Wang, Y.~Guo, K.~Lu, J.~Wan, D.~Wang, T.~Yao, and H.~Wu, ``Rcmp: Reconstructing rdma-based memory disaggregation via cxl,'' \emph{ACM Trans. Archit. Code Optim.}, vol.~21, no.~1, Jan. 2024.

\bibitem{rPCIeBench}
W.~Hou, J.~Zhang, Z.~Wang, and M.~Liu, ``Understanding routable {PCIe} performance for composable infrastructures,'' in \emph{21st USENIX Symposium on Networked Systems Design and Implementation (NSDI 24)}.\hskip 1em plus 0.5em minus 0.4em\relax Santa Clara, CA: USENIX Association, Apr. 2024, pp. 297--312.

\bibitem{ACaseAgainstCXL}
P.~Levis, K.~Lin, and A.~Tai, ``A case against cxl memory pooling,'' in \emph{Proceedings of the 22nd ACM Workshop on Hot Topics in Networks}, ser. HotNets '23.\hskip 1em plus 0.5em minus 0.4em\relax New York, NY, USA: Association for Computing Machinery, 2023, p. 18–24.

\bibitem{pond2}
D.~S. Berger, D.~Ernst, H.~Li, P.~Zardoshti, M.~Shah, S.~Rajadnya, S.~Lee, L.~Hsu, I.~Agarwal, M.~D. Hill, and R.~Bianchini, ``Design tradeoffs in cxl-based memory pools for public cloud platforms,'' \emph{IEEE Micro}, vol.~43, no.~2, p. 30–38, Mar. 2023.

\bibitem{jain2024memorysharingcxlhardware}
S.~Jain, N.~Yeleswarapu, H.~A. Maruf, and R.~Gupta, ``Memory sharing with cxl: Hardware and software design approaches,'' 2024.

\bibitem{TrEnv}
\BIBentryALTinterwordspacing
J.~Huang, M.~Zhang, T.~Ma, Z.~Liu, S.~Lin, K.~Chen, J.~Jiang, X.~Liao, Y.~Shan, N.~Zhang, M.~Lu, T.~Ma, H.~Gong, and Y.~Wu, ``Trenv: Transparently share serverless execution environments across different functions and nodes,'' in \emph{Proceedings of the ACM SIGOPS 30th Symposium on Operating Systems Principles}, ser. SOSP '24.\hskip 1em plus 0.5em minus 0.4em\relax New York, NY, USA: Association for Computing Machinery, 2024, p. 421–437. [Online]. Available: \url{https://doi.org/10.1145/3694715.3695967}
\BIBentrySTDinterwordspacing

\bibitem{HydraRPC}
T.~Ma, Z.~Liu, C.~Wei, J.~Huang, Y.~Zhuo, H.~Li, N.~Zhang, Y.~Guan, D.~Niu, M.~Zhang, and T.~Ma, ``{HydraRPC}: {RPC} in the {CXL} era,'' in \emph{2024 USENIX Annual Technical Conference (USENIX ATC 24)}.\hskip 1em plus 0.5em minus 0.4em\relax Santa Clara, CA: USENIX Association, Jul. 2024, pp. 387--395.

\bibitem{rpcool}
S.~Mahar, E.~Hajyjasini, S.~Lee, Z.~Zhang, M.~Shen, and S.~Swanson, ``Telepathic datacenters: Fast rpcs using shared cxl memory,'' 2024.

\bibitem{xu2024cxlsharedmemoryprogramming}
Y.~Xu, S.~Mahar, Z.~Liu, M.~Shen, and S.~Swanson, ``Cxl shared memory programming: Barely distributed and almost persistent,'' 2024.

\bibitem{liu2024dissectingcxlmemoryperformance}
\BIBentryALTinterwordspacing
J.~Liu, H.~Hadian, H.~Xu, D.~S. Berger, and H.~Li, ``Dissecting cxl memory performance at scale: Analysis, modeling, and optimization,'' 2024. [Online]. Available: \url{https://arxiv.org/abs/2409.14317}
\BIBentrySTDinterwordspacing

\bibitem{wang2024hitchhikersguideprogrammingoptimizing}
\BIBentryALTinterwordspacing
Z.~Wang, S.~Mahar, L.~Li, J.~Park, J.~Kim, T.~Michailidis, Y.~Pan, T.~Rosing, D.~Tullsen, S.~Swanson, K.~C. Ryoo, S.~Park, and J.~Zhao, ``The hitchhiker's guide to programming and optimizing cxl-based heterogeneous systems,'' 2024. [Online]. Available: \url{https://arxiv.org/abs/2411.02814}
\BIBentrySTDinterwordspacing

\bibitem{Asanović:EECS-2006-183}
K.~Asanović, R.~Bodik, B.~C. Catanzaro, J.~J. Gebis, P.~Husbands, K.~Keutzer, D.~A. Patterson, W.~L. Plishker, J.~Shalf, S.~W. Williams, and K.~A. Yelick, ``The landscape of parallel computing research: A view from berkeley,'' Tech. Rep. UCB/EECS-2006-183, Dec 2006.

\bibitem{YCSD}
B.~F. Cooper, A.~Silberstein, E.~Tam, R.~Ramakrishnan, and R.~Sears, ``Benchmarking cloud serving systems with ycsb,'' in \emph{Proceedings of the 1st ACM Symposium on Cloud Computing}, ser. SoCC '10.\hskip 1em plus 0.5em minus 0.4em\relax New York, NY, USA: Association for Computing Machinery, 2010, p. 143–154.

\bibitem{fio}
\BIBentryALTinterwordspacing
``Flexible i/o tester.'' 2024. [Online]. Available: \url{https://github.com/axboe/fio.}
\BIBentrySTDinterwordspacing

\bibitem{MERCI}
Y.~Lee, S.~H. Seo, H.~Choi, H.~U. Sul, S.~Kim, J.~W. Lee, and T.~J. Ham, ``\BIBforeignlanguage{en-US}{Merci: efficient embedding reduction on commodity hardware via sub-query memoization},'' in \emph{\BIBforeignlanguage{en-US}{Proceedings of the 26th ACM International Conference on Architectural Support for Programming Languages and Operating Systems}}, Apr 2021.

\bibitem{CXLINVIRTUAL}
\BIBentryALTinterwordspacing
Y.~Zhong, D.~S. Berger, C.~Waldspurger, R.~Wee, I.~Agarwal, R.~Agarwal, F.~Hady, K.~Kumar, M.~D. Hill, M.~Chowdhury, and A.~Cidon, ``Managing memory tiers with {CXL} in virtualized environments,'' in \emph{18th USENIX Symposium on Operating Systems Design and Implementation (OSDI 24)}.\hskip 1em plus 0.5em minus 0.4em\relax Santa Clara, CA: USENIX Association, Jul. 2024, pp. 37--56. [Online]. Available: \url{https://www.usenix.org/conference/osdi24/presentation/zhong-yuhong}
\BIBentrySTDinterwordspacing

\bibitem{cheng2024characterizingdilemmaperformanceindex}
\BIBentryALTinterwordspacing
R.~Cheng, Y.~Peng, X.~Wei, H.~Xie, R.~Chen, S.~Shen, and H.~Chen, ``Characterizing the dilemma of performance and index size in billion-scale vector search and breaking it with second-tier memory,'' 2024. [Online]. Available: \url{https://arxiv.org/abs/2405.03267}
\BIBentrySTDinterwordspacing

\bibitem{TensorOffloading}
D.~Xu, Y.~Feng, K.~Shin, D.~Kim, H.~Jeon, and D.~Li, ``Efficient tensor offloading for large deep-learning model training based on compute express link,'' in \emph{SC24: International Conference for High Performance Computing, Networking, Storage and Analysis}, 2024, pp. 1--18.

\bibitem{cxldb01}
M.~Ahn, A.~Chang, D.~Lee, J.~Gim, J.~Kim, J.~Jung, O.~Rebholz, V.~Pham, K.~Malladi, and Y.~S. Ki, ``Enabling cxl memory expansion for in-memory database management systems,'' in \emph{Proceedings of the 18th International Workshop on Data Management on New Hardware}, ser. DaMoN '22.\hskip 1em plus 0.5em minus 0.4em\relax New York, NY, USA: Association for Computing Machinery, 2022.

\bibitem{cxldb02}
\BIBentryALTinterwordspacing
A.~Lerner and G.~Alonso, ``Cxl and the return of scale-up database engines,'' \emph{Proc. VLDB Endow.}, vol.~17, no.~10, p. 2568–2575, Aug. 2024. [Online]. Available: \url{https://doi.org/10.14778/3675034.3675047}
\BIBentrySTDinterwordspacing

\bibitem{cxldb03}
N.~Riekenbrauck, M.~Weisgut, D.~Lindner, and T.~Rabl, ``A three-tier buffer manager integrating cxl device memory for database systems,'' in \emph{2024 IEEE 40th International Conference on Data Engineering Workshops (ICDEW)}, 2024, pp. 395--401.

\bibitem{Lee2024DatabaseKS}
\BIBentryALTinterwordspacing
S.~Lee, A.~Lerner, P.~Bonnet, and P.~Cudr{\'e}-Mauroux, ``Database kernels: Seamless integration of database systems and fast storage via cxl,'' in \emph{Conference on Innovative Data Systems Research}, 2024. [Online]. Available: \url{https://api.semanticscholar.org/CorpusID:266732447}
\BIBentrySTDinterwordspacing

\bibitem{cxlfass02}
D.~Boles, D.~Waddington, and D.~A. Roberts, ``Cxl-enabled enhanced memory functions,'' \emph{IEEE Micro}, vol.~43, no.~2, pp. 58--65, 2023.

\bibitem{cxlfass03}
A.~Patil, V.~Nagarajan, N.~Nikoleris, and N.~Oswald, ``Āpta: Fault-tolerant object-granular cxl disaggregated memory for accelerating faas,'' in \emph{2023 53rd Annual IEEE/IFIP International Conference on Dependable Systems and Networks (DSN)}, 2023, pp. 201--215.

\bibitem{li2023understandingoptimizingserverlessworkloads}
\BIBentryALTinterwordspacing
Y.~Li and S.~Yao, ``Understanding and optimizing serverless workloads in cxl-enabled tiered memory,'' 2023. [Online]. Available: \url{https://arxiv.org/abs/2309.01736}
\BIBentrySTDinterwordspacing

\bibitem{mpk}
\BIBentryALTinterwordspacing
J.~Corbet, ``Memory protection keys.'' 2015. [Online]. Available: \url{https://lwn.net/Articles/643797/}
\BIBentrySTDinterwordspacing

\bibitem{mutlu2023memorycentriccomputing}
\BIBentryALTinterwordspacing
O.~Mutlu, ``Memory-centric computing,'' 2023. [Online]. Available: \url{https://arxiv.org/abs/2305.20000}
\BIBentrySTDinterwordspacing

\bibitem{Mcc1}
A.~Gebregiorgis, H.~A. Du~Nguyen, J.~Yu, R.~Bishnoi, M.~Taouil, F.~Catthoor, and S.~Hamdioui, ``A survey on memory-centric computer architectures,'' \emph{J. Emerg. Technol. Comput. Syst.}, vol.~18, no.~4, Oct. 2022.

\end{thebibliography}
%


\end{document}